\documentclass[twocolumn]{aastex62}

\usepackage{amsmath} 
\usepackage{multirow}
\usepackage{hhline}
\usepackage{xcolor}
\usepackage{rotating}

\def\xc{\ensuremath{X_{\rm C}}}
\def\xci{\ensuremath{X_{\rm C}^{\rm i}}}
\def\xcm{\ensuremath{X_{\rm C}^{\rm m}}}
\def\xco{\ensuremath{X_{\rm C}^{\rm o}}}

\newcommand\Tstrut{\rule{0pt}{2.6ex}}       
\newcommand\Bstrut{\rule[-0.9ex]{0pt}{0pt}} 
\newcommand{\TBstrut}{\Tstrut\Bstrut} 

\shorttitle{Carbon as a diagnostic of explosion mechanisms}
\shortauthors{Heringer et al.}

\begin{document}

\title{Spectral sequences of Type Ia supernovae. II. Carbon as a diagnostic tool for explosion mechanisms}

\author{E. Heringer}
\affiliation{Department of Astronomy \& Astrophysics, University of Toronto, 50 Saint George Street, Toronto, ON, M5S 3H4, Canada}
\author{M. H. van Kerkwijk}
\affiliation{Department of Astronomy \& Astrophysics, University of Toronto, 50 Saint George Street, Toronto, ON, M5S 3H4, Canada}
\author{S. A. Sim}
\affiliation{Astrophysics Research Centre, School of Mathematics and Physics, Queen’s University Belfast, Belfast BT7 1NN, UK}
\author{W. E. Kerzendorf}
\affiliation{European Southern Observatory (ESO), Karl-Schwarzschild-Stra\ss{}e 2, D-85748 Garching, Germany}
\author{Melissa L. Graham}
\affiliation{Department of Astronomy, University of Washington, Box 351580, U.W., Seattle, WA, 98195, USA}

\begin{abstract}
How an otherwise inert carbon-oxygen white dwarf can be made to explode as a Type Ia supernova remains unknown. A promising test of theoretical models is to constrain the distribution of material that is left unburned, in particular of carbon. So far, most investigations used line identification codes to detect carbon in the ejecta, a method that cannot be readily compared against model predictions because it requires assumed opacities and temperatures. Here, we instead use tomographic techniques to investigate the amount of carbon in the inner layers of SN~2011fe, starting from the previously published tomographic analysis of \cite{Mazzali2014_tomography}. From the presence of the carbon feature in the optical at early epochs and its disappearance later on, we derive an average carbon mass fraction between 0.001 and 0.05 for velocities in the range $13500 \lesssim v \lesssim 16000\ \rm{km\ s^{-1}}$, and an upper limit of 0.005 inside that region. Based on our models and the assumed density profile, only small amounts of carbon should be in the neutral state, too little to be responsible for features seen in near-infrared spectra that were previously identified as due to neutral carbon; We discuss possible reasons for this discrepancy. We compare our results against a suite of explosion models, although uncertainties in both the models and our simulations make it difficult to draw definitive conclusions.
\end{abstract}


\section{Introduction}
\label{sec:introduction}

There is strong evidence that type Ia supernovae (SNe~Ia) originate from explosions of white dwarfs (WDs) in binary systems. However, both the explosion mechanism and the stellar evolutionary path remain unknown (for a review see \citealt{Livio2018_review}).

At least in part, the confusion surrounding the explosion mechanism arises from the fact that in a detonation the burning products are determined primarily by the local density of the fuel being consumed. For instance, carbon will remain mostly unburned at densities $\lesssim 10^{5}\ \rm{g\ cm^{-3}}$ \citep{Shigeyama1992_subChandra, Fink2010_helium}. In principle, models with different progenitor masses can still produce similarly plausible spectra provided that in the evolution up to the actual detonation the density profile of the star becomes similar. For instance, in the delayed detonation transition model (DDT, e.g. \citealt{Khokhlov1991_DDT, Blondin2013_DDT}), a near-Chandrasekhar WD expands during the deflagration phase; the detonation that follows then propagates across lower densities, producing a distribution of elements similar to that of a pure detonation of a sub-Chandrasekhar WD \citep{Sim2010_subChandra}.

In this series of papers we use the fast Monte Carlo radiative transfer code {\sc tardis} \citep{Kerzendorf2014_TARDIS} to try to determine whether the spectral features of goups of SNe~Ia can be explained as variations of a single physical parameter. In paper I \citep{Heringer2017_sequence}, we showed that  both normal and subluminous events can be reproduced using the same ejecta structure as long as the temperature was properly adjusted, suggesting that these events share a common explosion mechanism.

Here, we turn our attention to the distribution of carbon in the ejecta of SN~2011fe, a proto-typical normal event that was discovered possibly within a day of explosion, exhibited carbon and oxygen features in its pre-maximum spectra, and was extensively observed and analyzed (e.g., \citealt{Nugent2011_WD,Parrent2012_carbon,Vinko2012_11fe}). SN~2011fe reached its rest-frame $B$-band peak brightness on 2011 September 10 (MJD 55814.30 $\pm$ 0.06), corresponding to a time after explosion of $\sim$19 days \citep[][M14 hereafter]{Mazzali2014_tomography}. The exact rise time, however, will depend on how the $^{56}$Ni produced by the explosion is distributed in the ejecta--if there is little radioactive material near the surface, a ``dark phase" may follow before any light can escape (e.g. \citealt{Piro_darkphase}). M14 estimated this phase to last $\sim$1 day for SN~2011fe.

In principle, the abundance distribution of any given element could serve as a diagnostic tool for explosion mechanisms. However, carbon should be less susceptible to uncertainties pertaining to the precise nucleosynthesis, because the presence of unburned material is determined by a single burning process, compared to the multiple reactions required to form heavier elements. For instance, different groups have computed how much $^{56}$Ni is produced by the detonation of a WD of a given mass, finding discrepancies as large as a factor of 2, possibly due to differences in the nuclear reaction network adopted (see \citealt{Shen2018_ddet} and references therein).  Similarly, oxygen is less suitable because it is also a burning product of carbon (though in the nebular phase, which probes the deep interior, is very valuable: for instance, for SN~2010lp, an atypical SN~Ia, it was used to infer an asymmetric explosion; \citealt{Taubenberger2013_nebular}).

The usefulness of carbon as a diagnostic tool arises because theoretical models predict the presence of unburned material at widely different regions of the ejecta. For instance, the violent merger model of \citet{Pakmor2012_DD} leaves a large mass fraction ($\sim\!0.1$) of carbon mixed throughout the whole ejecta, while 3D simulations of the DDT scenario predict much more moderate amounts ($\sim\!0.01$) of carbon at velocities of $\sim 10000\ \rm{km\ s^{-1}}$ \citep{Seitenzahl2013_DDT}. In contrast, carbon might be concentrated in an intermediate layer in the double detonation scenario, as a result of the combustion of a helium shell \citep{Fink2010_helium, Sim2012_doubledet}. In models in which the supernova is approximated as the spontaneous detonation of a sub-Chandrasekhar WD with negligible helium mass, unburned material is expected only in the very outermost regions of the ejecta, at high velocities \citep{Shen2018_ddet}.

\citet{Wheeler1998_nir} were among the first to investigate carbon in SNe~Ia. They used a full radiative transport code \citep{Hoflich1995_radiative} to study SN~1986G and SN~1994D, finding the presence of Mg and O at similar layers in these events. Because Mg is a product of explosive burning of C, they were able to infer, based on delayed detonation models, that the transition between C and O burning takes place at $\sim 15000-16000\ \rm{km\ s^{-1}}$. \citet{Hoflich2002_nir} used the same code to study the subluminous SN~1999by, leading to the identification of a neutral carbon feature in the near-infrared (NIR) spectrum near maximum.

Subsequent research has used the parametrized spectral synthesis code \textsc{SYNOW} \citep{Branch2002_SYNOW} to identify carbon signatures in the optical spectra of individual SNe~Ia (SN~1998aq, \citealt{Branch2003_carbon}; SN~1999ac, \citealt{Garavini2005_carbon} and SN~2006D, \citealt{Thomas2007_carbon}). More recently, the SYNAPPS package (an automated version of \textsc{SYNOW}, \citealt{Thomas2011_synapps}) has been used to suggest the presence of unburned material in SN~2011fe at $\sim 10000-16000\ \rm{km\ s^{-1}}$ \citep{Parrent2012_carbon}.

Results from these semi-empirical studies include that the fraction of normal SNe~Ia that exhibit a carbon feature if observed early enough is $\sim\!30\%$, and possibly higher if in some cases the carbon feature is blended with other features due to high Doppler shifts \citep{Thomas2011_carbon, Parrent2011_carbon, Folatelli2012_carbon, Silverman2012_BSNIP_IV}.  Compared to normal SNe~Ia, there appears to be a paucity of subluminous and overluminous events that show carbon (\citealt{Parrent2011_carbon}; see \citealt{Chomiuk2016_subtypes} for a short summary of distinct subtypes of SNe~Ia). On the other hand, events that belong to the so called super-Chandrasekhar mass Ia subclass show a conspicuous carbon feature, significantly stronger than in normal events (e.g. \citealt{Hachinger2012_tomography}).

While parametrized packages, such as \textsc{SYNOW}, suffice to provide identifications and velocity estimates, they require temperature scales and reference opacities passed by hand and thus cannot be used to infer abundance ratios between elements, let alone quantitative abundance profiles across the ejecta. Detailed radiative transport codes (i.e. full NLTE or multi-dimensional) can be used to check abundance profiles, but their computational cost is too high to explore a large parameter space. In order to run the large number of simulations required to obtain abundance distributions of elements, one has to resort simplifications, and, e.g., use 1D codes with simplified excitation/ionization approximations.

Such simplified codes have been used for tomographic analyses, where the distribution of elements above the photosphere\footnote{Defined here as an inner boundary of the simulation domain where the radiation field is assumed to be that of a blackbody.} is estimated by fitting the spectra at different epochs. As the ejecta expand and the photosphere recedes, deeper layers of the ejecta become optically thin, allowing the inference of the distribution of various elements. Here, we follow this approach to estimate the mass fraction of carbon relatively deep in the ejecta of SN~2011fe ($7850 \lesssim v \lesssim 16000{\rm\,km\,s^{-1}}$) and compare it against model predictions. The study of the outermost layers could also potentially discriminate between models; However, for reasons we discuss in appendix \ref{sec:appendix}, we find that such analysis would require a more realistic tomographic analysis of the earliest spectrum, at $t_{\rm exp}$=3.7\,d. Redoing the tomography of SN~2011fe is beyond the scope of this work and therefore we reserve the analysis of the high velocity layers to an appendix and limit ourselves to place only conservative constraints on the carbon mass fraction in that zone.

Following paper I \citep{Heringer2017_sequence}, as a starting point we use the tomographic analysis of SN~2011fe by M14, which led to density and abundance profiles with which the observed spectra during the photospheric phase (up to one week after maximum) can be reproduced, at least qualitatively (so far, no code can reproduce observed spectra well enough for meaningful quantitative assessments).

From their tomographic analysis of SN~2011fe, M14 suggest that carbon is present (mass fraction $\sim\!0.01$) down to $\sim\!8000\ \mathrm{km\ s^{-1}}$.  This velocity is lower than predicted by many explosions models (see Section \ref{sec:results}) and also lower than typically inferred for other SN Ia. For instance, \citet{Folatelli2012_carbon} and \citet{Silverman2012_BSNIP_IV} derived typical Doppler velocities (measured from the absorption minimum)\footnote{Note that the Doppler velocity of the red wing of the carbon feature in the optical may extend to $\sim 8000\ \mathrm{km\ s^{-1}}$, but this does not imply that unburned material is present at such velocities, as further discussed in \S \ref{subsec:scan}.} of $\sim\!12000\ \mathrm{km\ s^{-1}}$, while \citet{Parrent2011_carbon} used \textsc{SYNOW} to estimate velocities, obtaining similar values for the core-normal \citep{Branch2006_subclasses} SNe\ Ia in their sample. Inspired by this hint of carbon at depth and the scarcity of reliable measurements of the carbon profile in SNe~Ia, we decided to re-assess the distribution of unburned material in SN~2011fe.

Focusing on a single element allows us to not just derive a best mass fraction, but a range of mass fractions that can describe the data, thus giving better hope of placing realistic constraints on the location and distribution of carbon that are useful for comparison against explosion models. It also helps to study more in depth the physical conditions that are relevant for the formation of the carbon feature (at least under the approximations we adopt), both in the optical and in the NIR, which may help to shed light on why the feature is more often seen in some subtypes than others.

This paper is divided as follows: in \S \ref{sec:methods} we describe our methodology and in \S \ref{sec:results} we constrain the carbon mass fraction in the ejecta of SN~2011fe, which is then compared against model predictions in \S \ref{sec:model_comparison}. In \S \ref{sec:trough} we discuss the physical conditions that are relevant for the formation of the singly-ionized carbon feature in the optical, while in \S \ref{sec:neutral} we focus on neutral carbon and its possible impact on the NIR part of the spectrum. In \S \ref{sec:uncertainties} we discuss possible sources of uncertainties and the limitations of our analysis. Our conclusions are presented in \S \ref{sec:ramifications}.  

\section{Methods}
\label{sec:methods}

To constrain the distribution of carbon in SN~2011fe, we use the spectral synthesis code TARDIS v1.5dev3181 \citep{Kerzendorf2014_TARDIS}. This code uses Monte Carlo radiative transfer through spherically symmetric ejecta.  It treats the densest parts of the ejecta as an optically thick region that produces a blackbody distribution of photons. The photons are represented by packets which can interact with the layers above either via line interaction (in the Sobolev approximation) or electron scattering. TARDIS follows the approach and is implemented similarly to the \cite{Lucy1999_code} code used by M14, except for possible differences in the atomic physics (see \S 3 of \citealt{Kerzendorf2014_TARDIS} for a detailed description), and generally produces consistent predicted spectra.

We chose TARDIS as a compromise between physical accuracy and computational cost in order to explore the parameter space of carbon mass fractions: opacities and temperatures across the ejecta are computed self-consistently, which is a clear improvement over line identification codes, but, as in the code used by M14, several approximations are made in order to reduce to computation time. The approximations that might affect our conclusions most are: \textit{(i)} The ejecta are assumed to have no energy deposition and thermalization processes above the photosphere. \textit{(ii)} The ejecta are treated as spherically symmetric, i.e., we simulate just the radial dimension\footnote{We usually take velocity as the radial coordinate, but this can readily be mapped to a radius at each given epoch since the ejecta are in homologous expansion at the timescales relevant for this work.} and cannot capture any possible line of sight effects if the explosion was asymmetric. \textit{(iii)} The distribution of ions and the population of atomic levels are computed using only a simple NTLE approximation, where these quantities are corrected by a ``dilution'' factor\footnote{We adopt the \texttt{nebular} and \texttt{dilute-lte} ionization and excitation modes in TARDIS, respectively.}.  We do not solve the full set of statistical equilibrium equations and thus ignore non-thermal processes (see \citealt{Mazzali1993_code} and \citealt{Kerzendorf2014_TARDIS} for detailed descriptions.) \textit{(iv)} Parameters other than the carbon abundance are fixed at the values found by M14, i.e., we do not attempt to see whether a change in the carbon feature due to a change in carbon abundance can be compensated by changes in the other parameters. Therefore, our results and conclusions are valid only within the scope of these approximations and comparison to models inherit these limitations.

\subsection{Input parameters}
\label{subsec:input_pars}

Like in Paper I \citep{Heringer2017_sequence}, we start from the tomography of SN~2011fe by M14.\footnote{Since Paper I, we have further improved our model inputs by interpolating the abundances provided in Fig.~10 of M14, leading to slightly different synthetic spectra. This, however, does not affect the conclusions presented in paper I.} We use the model parameters reported in their Table 6 and Fig.~10 and we generally find good agreement with the data (see Fig.~\ref{Fig:scan} below). Small discrepancies between the synthetic spectra are likely due to minor differences in the codes (and secondary input parameters). The only significant difference we find is for the 3.7 days spectrum, near $\lambda = 4000\, \text{\AA}$; this discrepancy is likely due to small abundance mismatches of Cr, Ti and/or Fe above the photosphere, located at $v_{\rm inner}=13300\ \mathrm{km\ s^{-1}}$, which can strongly influence the opacity at the blue end, even if the other parameters are similar (e.g., we infer a temperature of $10740$~K at the photosphere, which agrees with the value of $10800$~K reported by M14). Given their location, however, these abundance differences are unlikely to affect any of the conclusions we draw regarding carbon.

For the density profile, we use M14's ``$\rho$-11'' profile. At the velocities relevant for our work ($v\lesssim 19000{\rm\,km\,s^{-1}}$), this is similar to other profiles that were tested, viz., W7 of \cite{Nomoto1984_W7} and WDD1 of \cite{Iwamoto1999_WDD}, but differs at higher velocities, where it is intermediate between W7 and WDD1, leading to a better match to the UV part of early spectra. We later show in \S \ref{sec:model_comparison} that, in the velocities of interest, this density profile is similar to those of the explosion models we compare against.

The times after explosion ($t_{\rm{exp}}$) for the spectra used in this work are listed in Table \ref{tb:quantities} below, together with the corresponding velocities of the inner boundary in our simulations ($v_{\rm{inner}}$, also referred to as ``photospheric" velocity; these follow directly from M14). For comparison with our simulations, both the observed and simulated spectra were normalized by the mean flux computed in the region defined by $4000\, \text{\AA} \leq \lambda \leq 9000\, \text{\AA}$.

\section{The Distribution of Carbon in SN 2011fe}
\label{sec:results}

In order to constrain the distribution of carbon in SN~2011fe, we first determine a velocity scale $v_{\rm{lim}}$ -- and, thus, equivalently an enclosed mass -- below which carbon is not required to explain the data. Next, we abandon the M14 carbon distribution and instead approximate it assuming constant \xc\ in two regions of interest: immediately below and above $v_{\rm{lim}}$, aiming to determine how much carbon could be ``hidden'' below the velocity threshold and what is the minimum amount of carbon required above it. To this purpose, we inspect the most prominent carbon feature in the optical spectra, due to the \ion{C}{2} $\lambda$6580 line (e.g. \citealt{Folatelli2012_carbon}).

We note that in all cases, variations in the carbon mass fraction are made at the expense of the most abundant element in each layer so that the relative change of that element and the impact on the overall spectra are minimized.

\subsection{The depth down to which carbon has to be present in SN~2011fe}
\label{subsec:scan}

We start our analysis by revisiting the distribution of carbon in the ejecta of SN~2011fe inferred by M14 (see the bottom panel of their Fig.~10). Their adopted distribution includes carbon at velocities as low as $\sim\!8000\ \mathrm{km\ s^{-1}}$, comparable to the velocity of the inner boundary at maximum light, $v_{\rm inner}=7850\ \mathrm{km\ s^{-1}}$, suggesting the presence of unburned material relatively deep in the ejecta.

Since the carbon feature appears stronger in the synthetic spectra of M14 than seen in the data (see their Fig. 8 and 9, near $\lambda \approx 6300$\, \text{\AA}), we first aim to determine quantitatively the velocity $v_{\rm{lim}}$ below which carbon is not required to explain the data. For this purpose, we perform a ``depth scan,'' where we adopt the carbon profile of M14 but remove all carbon (i.e., set $\xc=0$) below a given velocity cut, $v_{\rm{cut}}$, which we vary in steps of $500\ \mathrm{km\ s^{-1}}$. $v_{\rm{lim}}$ is then determined as the largest $v_{\rm{cut}}$ for which the simulated spectra are consistent with the data.

\begin{figure}
\epsscale{1.1}
\plotone{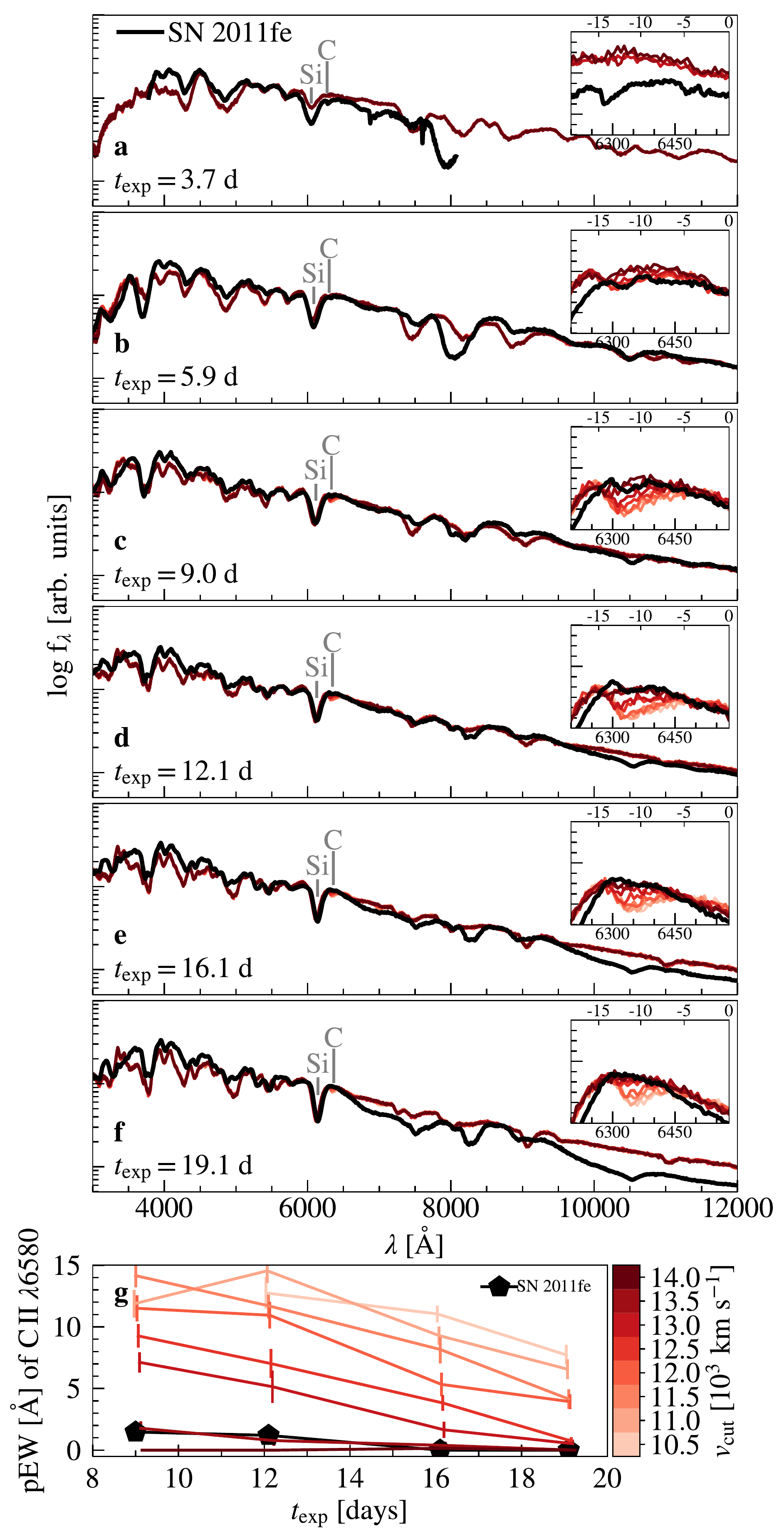}
\caption{The depth down to which carbon has to be present. {\em Panels \textbf{a}--\textbf{f}:\/} Observed SN~2011fe spectra (black lines) and simulated spectral sequences at $t_{\rm exp}$=3.7, 5.9, 9, 12.1, 16.1 and 19.1 days, respectively, with insets showing enlargements around the \ion{C}{2}$\lambda$6580 feature. The top $x$-axis of these insets also include a Doppler velocity scale, in units of $1000{\rm\,km\,s^{-1}}$.  {\em Panel \textbf{g}:\/} Observed and simulated pseudo-equivalent widths as a function of epoch, with observed values in black and simulated values color-coded by the velocity $v_{\rm{cut}}$ below which the mass fraction of carbon is set to zero (uncertainties are indicated, but are smaller than the symbols for the observed points). The pseudo-equivalent widths of the carbon feature at $t_{\rm exp}$=3.7, 5.9 days are not included as it is difficult to define the feature boundaries at these epochs. One sees that including carbon below $v\sim 13500{\rm\,km\,s^{-1}}$ at the levels adopted by M14 leads to predicted carbon features that are stronger than observed.}
\label{Fig:scan}
\end{figure}

Our simulations are shown in Fig. \ref{Fig:scan}, where the top panels (\textbf{a}--\textbf{f}) correspond to different epochs and the bottom panel (\textbf{g}) shows the corresponding pseudo equivalent widths (pEW\footnote{The routine we employ here to compute pEW is made publicly available alongside TARDIS.}, e.g. \citealt{Garavini2007_pEW}) of the carbon feature for each choice of $v_{\rm{cut}}$. The models generally provide a good fit and one can see that a choice of $\xc=0$ at $v\leq 13500\ \rm{km\,s^{-1}}$ suffices to explain the data. In fact, with the original carbon distribution below $13500{\rm\,km\,s^{-1}}$, for which $0.007 \lesssim \xc \lesssim 0.02$, the carbon features in the simulated spectra are stronger than those observed (while not providing a better fit anywhere else in the spectra). Based on the pEWs shown in panel \textbf{g}, we adopt $v_{\rm lim}=13500{\rm\,km\,s^{-1}}$.

We note that it is not clear why M14 included carbon at much lower velocities.  They refer to \citet{Parrent2012_carbon} and possibly they were motivated by the fact that the spectra presented in Fig.~4 of \citeauthor{Parrent2012_carbon} show the red wing of the \ion{C}{2} $\lambda$6580 feature extending down to $\sim\!8000\ \mathrm{km\ s^{-1}}$ in Doppler velocity.  The insets in Fig.~\ref{Fig:scan} help to illustrate why the red-most extent cannot be used directly to determine where carbon-rich layers are located in physical space. For instance, the models for which $v_{\rm{cut}}=10500{\rm\,km\,s^{-1}}$ show absorption down to Doppler shifts as low as $-5000{\rm\,km\,s^{-1}}$, even though no carbon is actually present at those depths. This is because the line-of-sight component of the velocity of parts of the ejecta that scatter a line can be smaller than the true velocity, i.e., the absorption seen at low velocity can be due to parts of the ejecta that are at higher velocity but are not directly moving towards us.

\subsection{Constraining the carbon profile}
\label{subsec:plateaus}

In order to constrain the range of carbon mass fractions that can reproduce the data, we next aim to determine for what \xc\ in a given velocity range the trough has a depth roughly consistent with the observations. Here, since the simulations are not yet at a level where they can reproduce spectra in detail, we refrain from employing any particular quantitative metric of consistency. Instead, we aim to determine upper and lower limits to the carbon mass fractions beyond which the simulated features are in obvious disagreement with the observed ones.

Since we compare with a feature that has only few parameters, it is clear that only limited information on the carbon profile can be gleaned. Indeed, we find that a simple two-zone model suffices, where we divide the velocity range into inner and middle regions, which are separated at $v_{\rm lim}=13500{\rm\,km\,s^{-1}}$, and which each have a constant mass fraction: \xci\ at $7850 \lesssim v \lesssim 13500{\rm\,km\,s^{-1}}$ and \xcm\ at $13500 \lesssim v \lesssim 16000{\rm\,km\,s^{-1}}$. Here, the lower velocity limit of the inner region is set by the photospheric velocity of our latest spectrum, while the upper limit of the middle region is somewhat arbitrary. We explore the effects of moving the boundary between the inner and middle regions in \S \ref{sec:uncertainties}, and discuss a trial in which we include an outer zone in appendix \ref{sec:appendix}.

In Fig.~\ref{Fig:plateaus} we compare simulated spectra for which the carbon mass fraction is varied in the inner and middle regions with observed carbon profiles. For each epoch, we run simulations for combinations of \xci, \xcm\ in the range $0 \leq \xci \leq 0.05$ and $0 \leq \xcm \leq 0.05$.  This choice of range covers both extremes, resulting in spectra that either contain too little carbon for the feature to manifest or too much carbon, producing a flux depression much stronger than seen in the data. Note that for all but the last epoch, the photosphere is outside the lower velocity limit of our inner region. Hence, with increasing time, one constrains the mass fraction at increasing depth, thus justifying the choice of our inner region being coarser ($\Delta v \sim 5500{\rm\,km\,s^{-1}}$) than the middle region ($\Delta v \sim 2500{\rm\,km\,s^{-1}}$).

\begin{figure*}
\epsscale{1.1}
\plotone{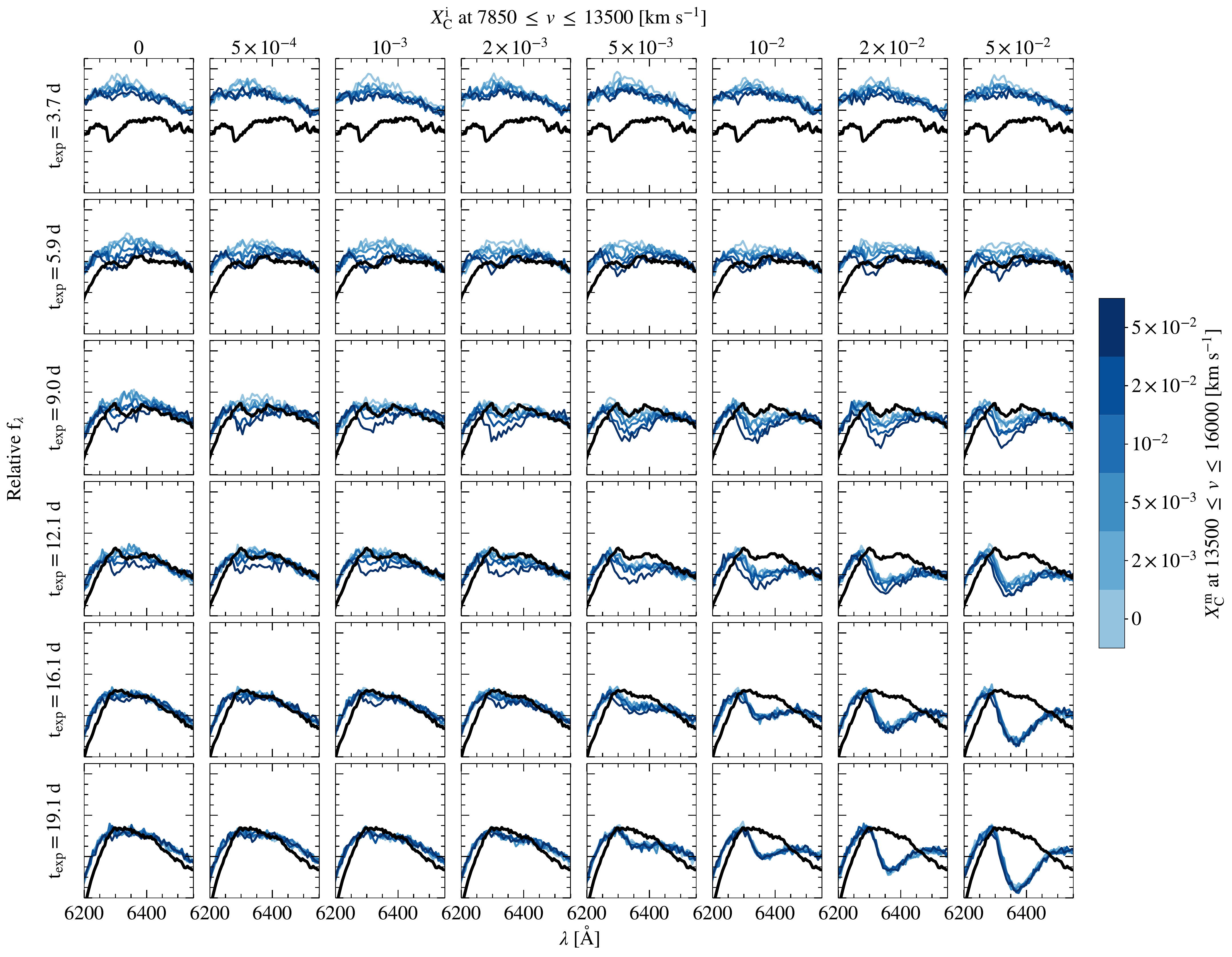}
\caption{Comparison of observed (black) and simulated (color) carbon profiles for a range of carbon abundances.  Rows correspond to different epochs, while columns map the carbon mass fraction \xci\ adopted for the inner region (velocity range $7850 \lesssim v \lesssim 13500{\rm\,km\,s^{-1}}$), and colors map the mass fraction \xcm\ in the middle region ($13500 \lesssim v \lesssim 16000{\rm\,km\,s^{-1}}$). One sees, for instance, that models with $\xci \geq 0.005$ cannot explain the data at $t_{\rm exp}$=16.1 and 19.1\,d for any valid choice of \xcm.}
\label{Fig:plateaus}
\end{figure*}

Inspecting Fig.~\ref{Fig:plateaus}, we infer for the middle region an upper limit of $\xcm<0.05$: for larger values the carbon feature is stronger than observed for any choice of \xci\ at $t_{\rm exp}=9\,$d.  For the inner region, the carbon fraction is constrained more effectively by the later epochs, since more of the region is exposed. Hence, for $t_{\rm exp} = 9$ days, we find $\xci<0.01$, while for later epochs the carbon feature is overpredicted unless $\xci < 0.005$.

Setting a lower limit to \xcm\ is more complicated because a deficit of carbon in the middle region may be compensated by larger fractions of carbon in an outer region. From Fig.~\ref{Fig:plateaus} alone, one would be tempted to set a lower limit of $\xcm>0.005$ based on the strength of the carbon feature at $t_{\rm exp}=5.9\ \rm{d}$. However, we show in Fig.~\ref{Fig:early_constraints} in appendix \ref{sec:appendix} that if we allow the carbon fraction outside this region to be larger, somewhat lower values may work. Based on the spectra at $t_{\rm exp}=5.9\,$d, we infer a more conservative lower limit of $\xcm \geq 0.001$.

Here, we note that inferences are hindered by the red wing of the Si feature not being predicted correctly, i.e. being offset by $\sim 50$\, \text{\AA} (the same can be seen in M14). As a consequence, the emission portion of this Si feature (i.e. a ``limb brightening'' effect [\citealt{Hoeflich1990_brightening}]) will also be somewhat shifted, thus affecting the precise ``pseudo-continuum''  across the wavelength region relevant for the \ion{C}{2}$\lambda$6580 line. Looking at the choice of $\xci\ = \xcm\ = 0$ in Fig \ref{Fig:plateaus}, we expect a shift in the pseudo continuum to have a minor effect for a qualitatively analysis like ours, especially at $t_{\rm exp}\geq 9.0\,$d. Thus, the poor fit of the Si feature should not affect our analysis of the carbon fraction at greater depth, but will be important for any future assessment of carbon at high velocities.

\begin{figure}
\epsscale{1.15}
\plotone{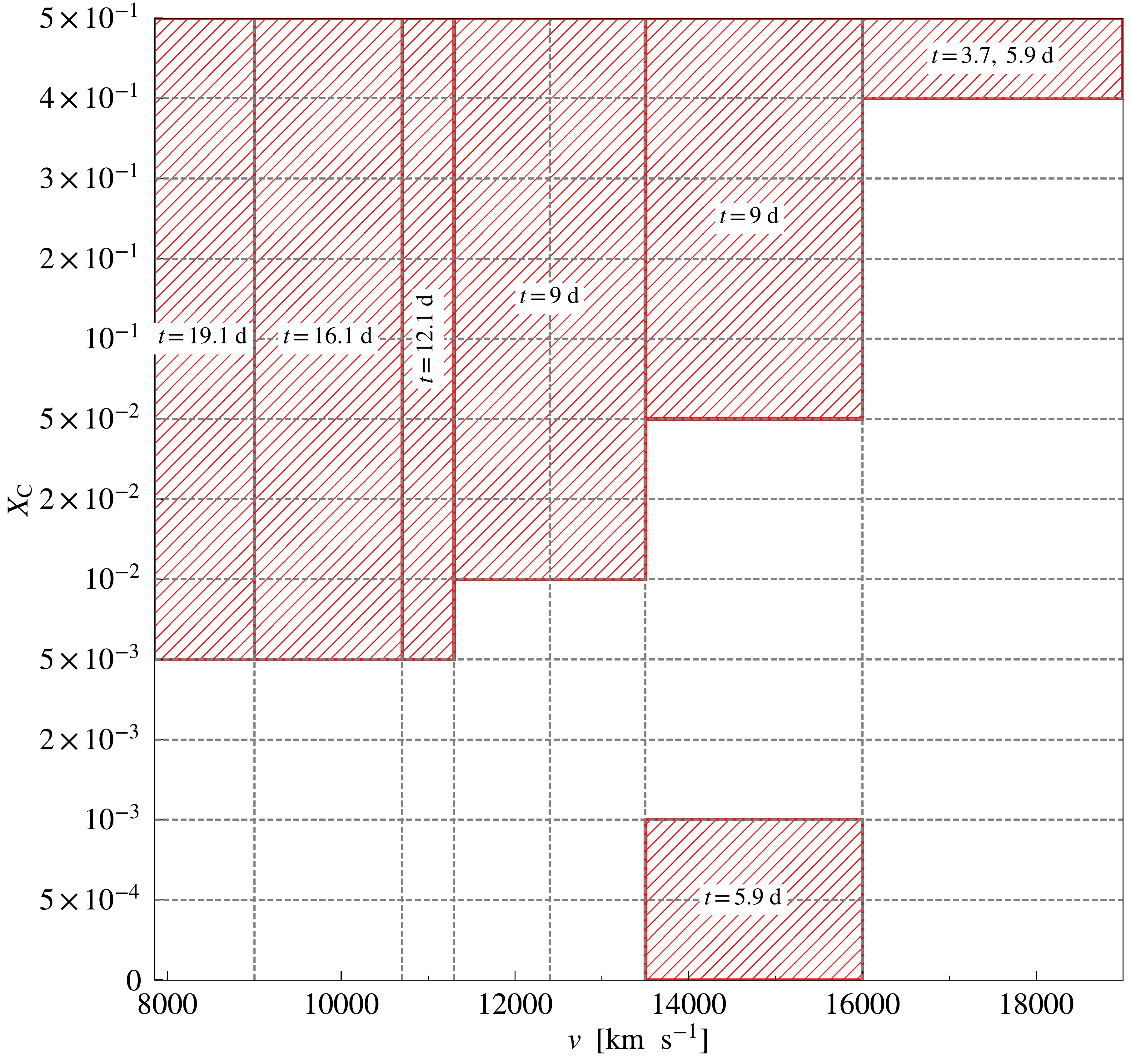}
\caption{Schematic plot showing the region of the carbon mass fraction parameter space that can be ruled out (hatched red) by comparing the spectra from a suite of simulations and the observed data (as shown in Fig. \ref{Fig:plateaus}.) Each hatched region is labeled according to the epoch relevant for its exclusion.}
\label{Fig:constraints}
\end{figure}

In Fig. \ref{Fig:constraints}, we indicate schematically the parts of the carbon mass fraction parameter space that we believe are ruled out, labeling each region with the epoch based on which we excluded it. These are meant to be conservative limits. In particular, not all combinations of $(\xci, \xcm)$ between their respective limits will produce spectra that match the data. Rather, the limits for one region indicate that there exists a choice of carbon fraction for the other region for which the data can be understood.

The above abundances are strictly for the adopted density profile.  Since the optical depth scales, to first order, with density, the corresponding mass of carbon $M_{\rm C}$ is a less model-dependent quantity.  For the middle region ($13500 \lesssim v \lesssim 16000{\rm\,km\,s^{-1}}$), we find that $10^{-4}< M_{\rm C} < 5\times 10^{-3}\,M_\odot$ is required to form the carbon feature.
In the inner region ($7850 \lesssim v \lesssim 13500{\rm\,km\,s^{-1}}$), we find an upper limit of $M_{\rm C} < 2.5\times 10^{-3}\,M_\odot$.

Finally, we note that carbon is predominantly singly ionized in the middle region, so uncertainties in the degree of ionization should be small and our inferred range of \xcm\ reasonably reliable. For the inner region, a significant fraction of carbon can be doubly ionized, although the singly-ionized fraction never becomes small. In Table \ref{tb:quantities}, we list inferred masses, ionization fractions and optical depths for reference values of $\xci=0.002$ and $\xcm=0.01$.

\begin{deluxetable}{lll}
\tablecaption{Summary of models. \label{tb:models}}
\tablehead{
Model & Scenario & References\TBstrut
}
\startdata
W7\dotfill & Fast deflagration & \citet{Nomoto1984_W7}\\[.8ex]
N100\dotfill & Delayed det. & \citet{Seitenzahl2013_DDT}\\ 
& transition & \citet{Ropke2012_N100} \\ 
& & \citet{Sim2013_radtransfer}\\[.8ex]
GCD200\dotfill & Gravitationally& \citet{Seitenzahl2016_gcd}\\
 & confined det. & \\[.8ex]
(1.1, 0.9)$\,M_\odot\ldots$ & Violent Merger & \citet{Pakmor2012_DD}\\[.8ex] 
$1.0\,M_{\odot}$\dotfill & Double det. & \citet{Fink2010_helium}\\[.8ex]
$1.0\,M_{\odot}$\dotfill & Spontaneous det. & \citet{Shen2018_ddet}
\enddata
\end{deluxetable}

\begin{figure*}
\epsscale{1.0}
\plotone{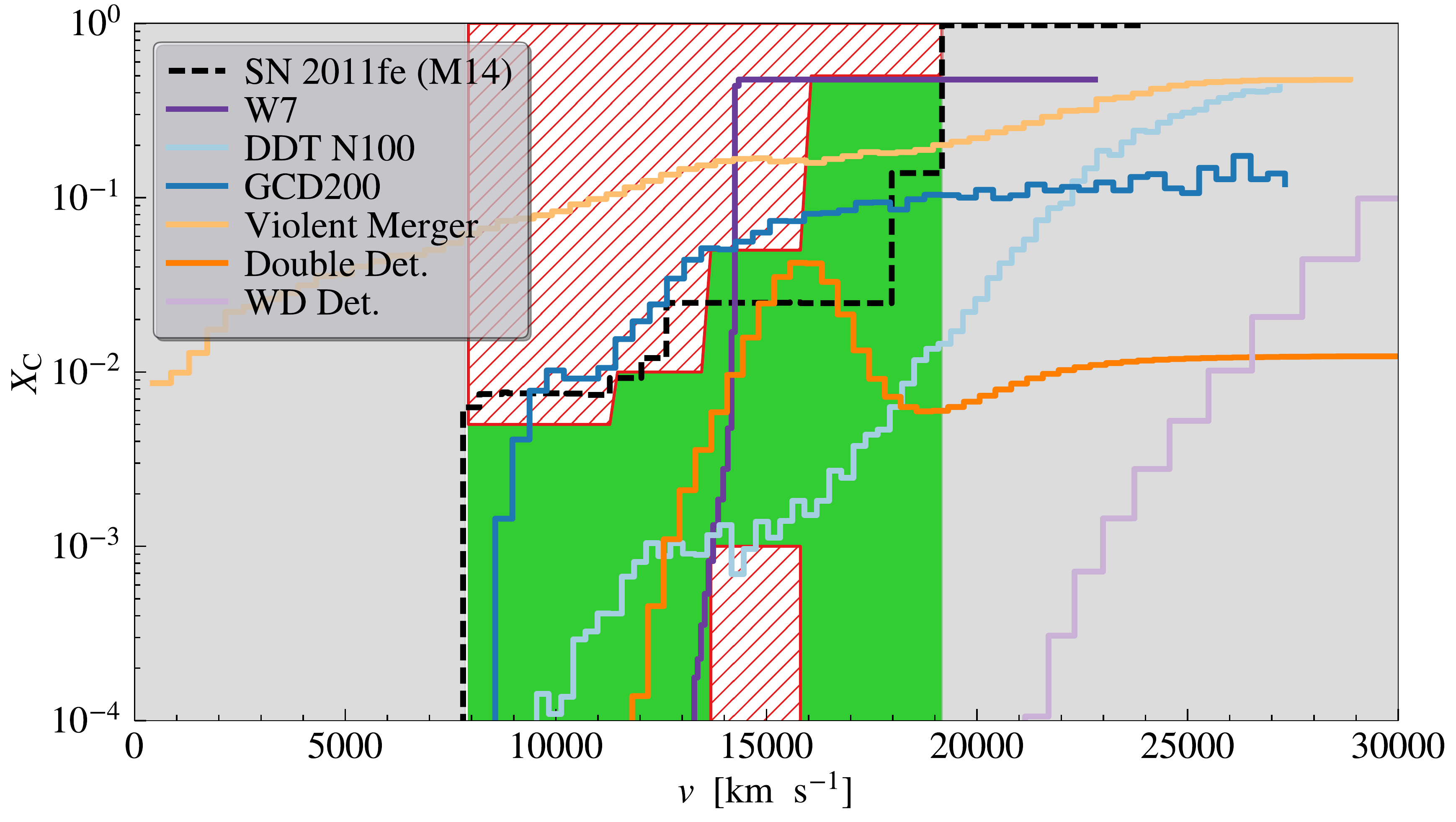}
\caption{Comparison of the allowed range of carbon mass fractions derived for SN~2011fe against explosion models. Background colors indicate that the region is ruled-out (hatched red), allowed (green) or unconstrained (gray), based on the models shown in Fig. \ref{Fig:plateaus} and \ref{Fig:early_constraints}, which are summarized in Fig. \ref{Fig:constraints}. For reference, the black line shows the carbon profile interpreted from the tomography analysis of M14, while the other lines represent the prediction from a suite of explosion models available in the literature. The respective references for each model can be found in Table \ref{tb:models}.}
\label{Fig:comparison}
\end{figure*}

\begin{figure}
\epsscale{1.2}
\plotone{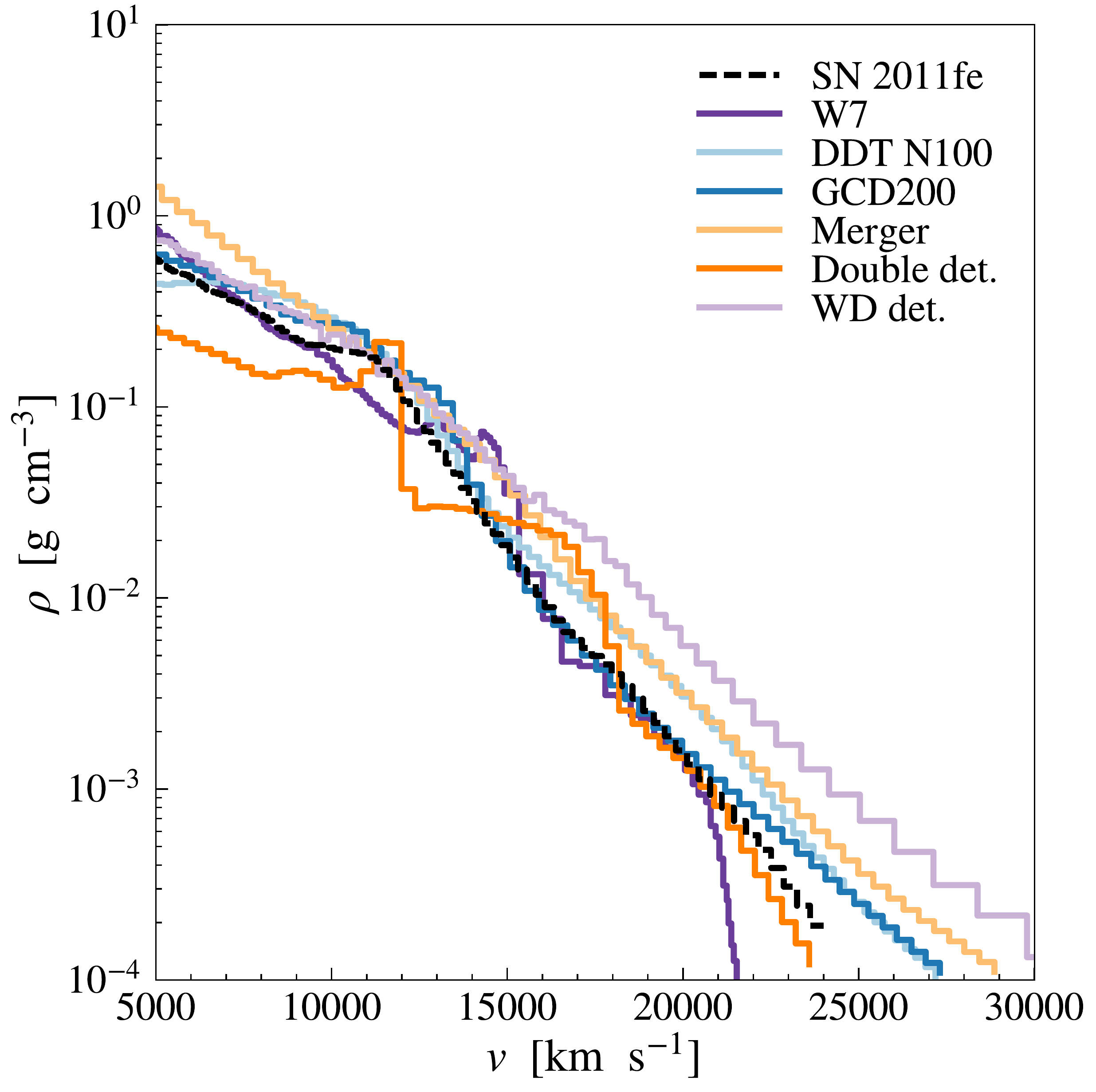}
\caption{Density profile at $100\ \rm{s}$ after the explosion for a suite of explosion models. The appropriate references for each model are given in Table \ref{tb:models}.}
\label{Fig:density}
\end{figure}

\section{Comparison to Explosion Models}
\label{sec:model_comparison}

We compare our constraints on the carbon mass fraction distribution with the predictions from different explosion models\footnote{Density and carbon abundance distributions for the explosion models were retrieved from HESMA (\citealt{Kromer2017_hesma}; \url{https://hesma.h-its.org}).}, as summarized in Table \ref{tb:models}. We show the comparison graphically in Fig.~\ref{Fig:comparison} and discuss the individual models below, but note here that, naturally, they predict not only distinct abundance profiles but also different density profiles, which are equally important in determining whether a feature manifests or not. Fortunately, as can be seen in Fig.~\ref{Fig:density}, the density profiles predicted by different models are remarkably homogeneous: they agree within a factor of 5 for $v\lesssim 19000{\rm\,km\,s^{-1}}$ and are even closer in our regions of interest (i.e. $7850 \lesssim v \lesssim 16000{\rm\,km\,s^{-1}}$). Indeed, we find that our conclusions below are unchanged if we compare predictions in terms of carbon mass (which is arguably more logical for line strengths, but we opted to show mass fraction instead as it more closely reflects the properties of the exploding white dwarf).

\subsection{Fast Deflagration (W7)}
\label{subsec:W7}

The W7 model \citep{Nomoto1984_W7} was part of a suite of spherically symmetric pure deflagration models. For W7, the predictions are a good match to observations, but it required a flame propagation speed much higher than expected theoretically or found in more recent pure deflagration models (which also produce significantly different outcomes; e.g., \citealt{Fink2014_def}). Nonetheless, observables derived from the W7 model remain a standard comparison baseline, and thus we include it for reference.

From  Figure \ref{Fig:comparison}, one sees that W7 predicts a sharp transition between regions where carbon is fully burned and not burned at all.  This is consistent with no carbon in our inner region, but seems to overpredict the carbon in the middle region. Indeed, we find a carbon feature that is too broad and blue if we run a simulation with a W7-like carbon distribution.

\subsection{Delayed Detonation (N100)}
\label{subsec:N100}

In delayed detonation models, run-away carbon fusion in a near-Chandrasekhar mass WD at first leads to a deflagration, which heats up and expands the white dwarf before transitioning to a detonation. These models are among the best studied and provide fairly good matches for observations.

For our comparison, we used data from the ``N100'' model \citep{Seitenzahl2013_DDT}, which was computed as part of a sequence of 3D simulations of delayed detonations, where the explosion strength is parametrized by the number of ignition kernels. The deflagration to detonation transition occurs when a pre-defined condition based on the local turbulent velocity is met (for a thorough description, see \citealt{Ciaraldi2013A&ADTD}).

In this model, the input parameters are such that the predicted peak brightness and spectra resemble those of normal SNe~Ia. The synthetic light curve for this model was computed by \citet{Sim2013_radtransfer} and compared to the normal SN~2005cf \citep{Pastorello2005_05cf}, exhibiting a reasonable agreement and only a modest dependency with the viewing angle, which is particularly small at the wavelengths relevant for the carbon feature. Some challenges to this model include the peak $B-V$ color being too red and the post maximum decline rate of the light curve being too fast when compared to typical values of normal SNe~Ia.

In general, the DDT models of \citet{Seitenzahl2013_DDT} predicted optical colors that were systematically redder than expected at peak and did not reproduce the width-luminosity relation \citep{Phillips1993_relation} in its entirety. One significant difference of this model compared to 1D DDT models (where the flame transition to detonation is typically assumed to occur at a density threshold; e.g., \citealt{Blondin2013_DDT}) is that, because of the buoyancy of the hot ignition kernels, the deflagration burning products are not necessary centrally located at low velocities and might be mixed through zones rich in intermediate mass elements, such as Si.

We find that the carbon mass fraction predicted, on average, by this model is formally within the allowed region we derive here. We note, however, that this model has relatively little carbon in both the middle and outer regions. In Fig. \ref{Fig:early_constraints} we see that $\xcm\ \sim 0.001$ requires $\xco\ \gtrsim 0.2$ for a clear carbon trough to be formed at $t_{\rm exp}=$ 5.9 and 9\,d, but N100 exhibits $\xco\ \lesssim 0.1$. We thus consider N100 to be a borderline case.

\subsection{Gravitationally Confined Detonation (GCD200)}
\label{subsec:gcd200}

A gravitationally confined detonation (GCD) might occur if, once a hot deflagration bubble breaks out of the surface, some or most of the material were to remain bound.  This could then wrap around the star, and trigger an off-center detonation upon convergence at the opposite side \citep{Plewa2004_GCD}. A general problem with these models is that they overpredict the abundances of iron-group elements at high velocities \citep{Seitenzahl2016_gcd}.

Nevertheless comparing our results to the simulations of \citet{Seitenzahl2016_gcd}, we find that the carbon abundance distribution is roughly consistent, if perhaps a little too high, especially around $v\simeq11000{\rm\,km\,s^{-1}}$. Given the approximations in our modeling, we also consider this to be a borderline case which we cannot rule out with confidence. We caution, though, that \citeauthor{Seitenzahl2016_gcd} note that their model is too bright to be representative of normal SNe~Ia.

\subsection{Violent Merger (1.1M$_\odot$+0.9M$_\odot$)}
\label{subsec:merger}

Some SNe~Ia might arise from the ``violent'' merger of two WDs, where the explosion is triggered at hot spots during the merging process. Existing studies for this scenario are not currently considered likely to be responsible for normal SNe~Ia, as it predicts, e.g., far stronger polarization that is observed \citep{Maund2013_VMpol, Bulla2016_VMpol}.

We find that it also is inconsistent with our constraints on the carbon abundance profiles, producing, on average, too much unburned material mixed throughout the ejecta.

\subsection{Double Detonation}
\label{subsec:doubledet}

In the double detonation scenario, a SN~Ia arises when a detonation in a surface helium shell triggers a secondary detonation in the WD core \citep{Livne1990_doubledet}. One of the outstanding questions for this class of models is whether the elements produced in the helium detonation would produce signatures, e.g., of iron-group elements, that are not observed \citep{Kromer2010_doubledet}. 

\citet{Fink2010_helium} calculated element distributions for a range of progenitor masses, under the assumption that each had a helium layer with the minimum mass required to trigger a detonation.  An interesting feature in these simulations is that unburned material is mostly confined to an intermediate layer -- the carbon in the outer layers is partially burned by the initial shell detonation. The velocity of this intermediate layer increases systematically with the strength of the explosion, because for the more massive white dwarfs that lead to brighter explosions, a smaller minimum helium mass is required, which leads to unburned carbon further out in the ejecta.

For our comparison, we used model~3 of \citet{Fink2010_helium} -- which corresponds to the double detonation of a 1.0~M$_\odot$ WD surrounded by a 0.055~M$_\odot$ He shell -- as it has the predicted $^{56}$Ni masses closest to the mass of $\sim\!0.45\,M_\odot$ estimated for SN~2011fe \citep{Nugent2011_WD}.  We find that the predicted distribution of carbon agrees reasonably well with the allowed region we infer in this work, especially considering that the predicted position of this layer may vary at the $1000{\rm\,km\,s^{-1}}$ level due to viewing angle effects. 

\subsection{Detonation of a Cold White Dwarf}
\label{subsec:WDdet}

Simulations of the detonation of cold WDs, i.e., with a detonation started in the centre without assigning a specific cause, have been shown to provide remarkably good matches to observations of SN~Ia \citep{Shigeyama1992_subChandra, Sim2010_subChandra}. These models could be seen as examples of the double detonation model in the limit where the helium shell can be disregarded, perhaps representative of, e.g., models in which the helium is accreted very rapidly, at the onset of a merger, leading to helium masses of $\lesssim\!10^{-2}\,M_\odot$ \citep{Shen2014_ddet}.   

For this class of models, the density gradient on the outside is steep and one thus expects carbon only at very high velocity. Indeed, in the recent simulations of detonations of cold WDs by \citet{Shen2018_ddet}, carbon is virtually absent at the velocities we cover: \xc\ $<10^{-4}$ for $v<21000\ \rm{km\ s^{-1}}$. \citeauthor{Shen2018_ddet} note, however, that a proper simulation of a double detonation would have to take into account the inward shock launched by the helium shell detonation, which might tamp the explosion and reduce the velocities of the outer layers. 

\subsection{Summary of the Model Comparisons}
\label{subsec:overall}

The carbon distributions predicted by the violent merger model of \citet{Pakmor2012_DD} and by the detonation of cold WDs model of \citet{Shen2018_ddet} both seem inconsistent with the constraints on the carbon mass fractions (and masses) we derived for SN~2011fe, with the former predicting, on average, too much and the latter too little unburned material at intermediate velocities (see Fig.~\ref{Fig:comparison}). We suggest that these differences are large enough for these models to be ruled out, despite the approximations involved in computing our synthetic spectra.

The carbon distributions predicted by the gravitationally confined detonation model of \citet{Seitenzahl2016_gcd} and by the N100 delayed detonation model of \citet{Seitenzahl2013_DDT} are roughly consistent with our results. This might provide a clue to what physical processes are needed to explain carbon distribution in the ejecta: in both mechanisms, the initial deflagration leads to relatively low density material at the locations that end up at the intermediate velocities where we infer carbon is present.

If the carbon distribution is indeed a consequence of the deflagration phase, then one might expect it to be somewhat stochastic and dependent on viewing geometry, thus providing a possible explanation for the variation in strengths of the carbon features in SN~Ia spectra \citep{Parrent2011_carbon}. This hypothesis, however, needs to be further investigated. For instance, \citet{Sim2013_radtransfer} showed that the light curves and spectra derived for the N100 model had only a mild dependency on viewing angle, despite significant anisotropies in the flame propagation. A similar conclusion was reached by \citet{Fink2018_diffrot}, who computed their observables in the context of DDTs in differentially rotating WDs.

Finally, the predicted carbon distribution for the double detonation model is also consistent with our results and thus cannot be ruled out.

\section{The Physical Conditions Relevant for the Carbon Feature}
\label{sec:trough}

In this section, we attempt to answer the following questions: \textit{Where in the ejecta is the C feature formed?}; \textit{Why does the C feature usually disappear near maximum light? \textit{and How much carbon can be hidden in normal SNe~Ia that do not exhibit a C feature?}} We will do this in the context of our models; uncertainties in those, including the NLTE approximations made in TARDIS, are addressed in Section~\ref{sec:uncertainties}. Unless otherwise stated, we will compare with a fiducial model that reproduces the observed features of SN~2011fe well, with $\xci=0.002$, $\xcm=0.01$ (and the carbon profile of M14 outside).

The carbon $\lambda6580$ feature arises from the $\rm{2s^23s}~^2{\rm S} \rightarrow \rm{2s^23p}~^2{\rm P}$ transition in \ion{C}{2} \citep{NIST_ASD}.
Because the ejecta are expanding, one expects a P-cygni profile.  This will be superposed on the red wing of the main Si~$\lambda6355$ feature, which means that, depending on the latter's strength, one may note the presence of carbon either through an actual trough or just through flattening of the silicon feature (as can be seen in Fig.~\ref{Fig:plateaus} for the different choices of \xcm\ at, e.g., $t_{\rm exp}=5.9\,$d and $\xci=0$).

\begin{figure}
\epsscale{1.15}
\plotone{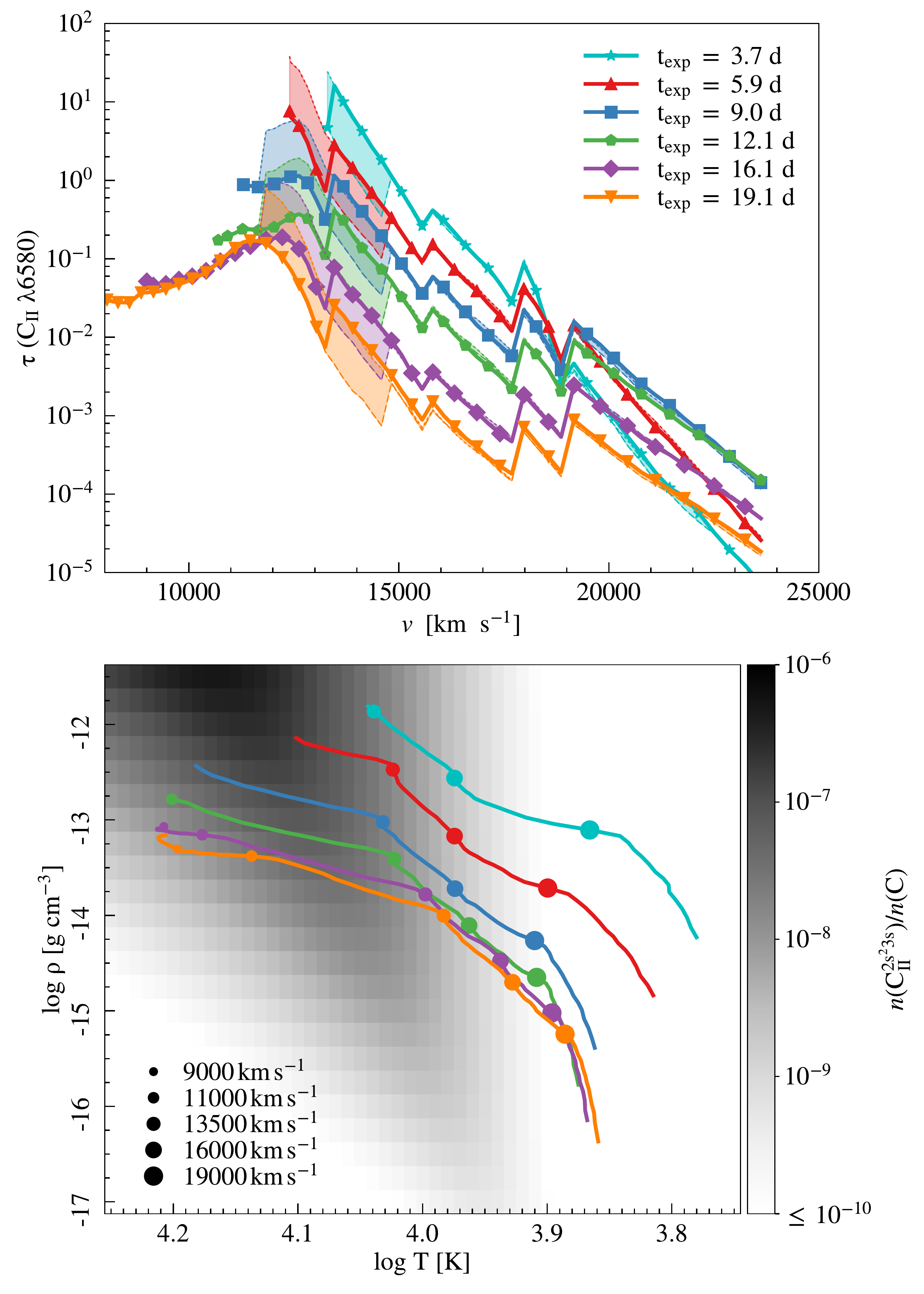}
\caption{\textit{Top:\/} Sobolev optical depth of the \ion{C}{2} $\lambda$6580 line in the fiducial models of SN~2011fe, with $\xci=0.002$ and $\xcm=0.01$, with different colors representing different epochs, as indicated. The shaded regions show the change in opacity if the velocity boundary is set to $12000\ {\rm\,km\,s^{-1}}$ (dotted line) or $15000\ {\rm\,km\,s^{-1}}$ (dashed line). The general decline reflects dilution of the column density with time due to the expansion of the ejecta, the jump in opacity corresponds to the change in carbon abundance between our inner and middle regions, and the decrease towards low velocities the increasing presence of \ion{C}{3}; \textit{Bottom:\/} Fraction of \ion{C}{2} in the $\rm{2s^23s}$ level as a function of temperature and density, as expected under the NLTE approximations adopted here. Overdrawn are temperature-density profiles for our fiducial SN~2011fe models, color-coded as in the top panel, and with circles marking, from right to left, the zones at $19000$, $16000$, $13500$, $11000$ and $9000\ {\rm\,km\,s^{-1}}$; note that the lowest velocities are below the inner boundary at the earliest epochs. One sees that while the fractional abundance of \ion{C}{2} at the $\rm{2s^23s}$ level is higher in the inner regions, the opacity is larger a few $1000\ {\rm\,km\,s^{-1}}$ above the ``pseudo-photosphere'', where the mass fraction of \ion{C}{2} is larger.}
\label{Fig:physics}
\end{figure}

\begin{deluxetable*}{lllllll}
\tablecaption{Relevant quantities. \label{tb:quantities}}
\tablecolumns{7}
\tablehead{\colhead{}& \multicolumn{6}{c}{\dotfill Epoch\tablenotemark{b} (d)\dotfill}\\
\colhead{Quantity\tablenotemark{a}} & \colhead{3.7} & \colhead{5.9} & \colhead{9} & \colhead{12.1} & \colhead{16.1} & \colhead{19.1}}
\startdata
\sidehead{Global properties}
$\log_{10} L/L_\odot$\dotfill & 7.903& 8.505& 9.041& 9.362& 9.505& 9.544\\
$v_{\rm inner}~({\rm km\,s^{-1}})$\dotfill& 13300& 12400& 11300& 10700& 9000& 7850\\
\sidehead{Inner part ($v_{\rm inner} < v \leq 13500{\rm\,km\,s^{-1}}$)}
$M_{\rm tot}~(M_\odot)$\dotfill & $0.0205$ & $0.0855$ & $0.2127$ & $0.2984$ & $0.5132$ & $0.6508$ \\
$M_{\rm C}~(M_\odot)$\dotfill & $0.0001$ & $0.0003$ & $0.0005$ & $0.0007$ & $0.0011$ & $0.0014$ \\
$M_{\rm C\, I}/M_{\rm C}$\dotfill & $0.000006$ & $0.000002$ & $0.000000$ & $0.000000$ & $0.000000$ & $0.000000$ \\
$M_{\rm C\, II}/M_{\rm C}$\dotfill & $0.9780$ & $0.8432$ & $0.3692$ & $0.2652$ & $0.2153$ & $0.2359$ \\
max $\tau(\mbox{\ion{C}{1}}~\lambda10693)$\ldots & $0.29421$ & $0.03836$ & $0.00426$ & $0.00111$ & $0.00041$ & $0.00022$ \\
max $\tau(\mbox{\ion{C}{2}}~\lambda6580)$\dotfill & $16.25423$ & $7.64200$ & $1.15749$ & $0.41434$ & $0.18804$ & $0.17041$ \\
\sidehead{Middle part ($13500 < v \leq 16000{\rm\,km\,s^{-1}}$)}
$M_{\rm tot}~(M_\odot)$\dotfill & $0.0683$ & $0.0683$ & $0.0683$ & $0.0682$ & $0.0682$ & $0.0682$ \\
$M_{\rm C}~(M_\odot)$\dotfill & $0.0007$ & $0.0007$ & $0.0007$ & $0.0007$ & $0.0007$ & $0.0007$ \\
$M_{\rm C\, I}/M_{\rm C}$\dotfill & $0.000015$ & $0.000005$ & $0.000002$ & $0.000001$ & $0.000001$ & $0.000001$ \\
$M_{\rm C\, II}/M_{\rm C}$\dotfill & $0.9953$ & $0.9929$ & $0.9778$ & $0.9751$ & $0.9932$ & $0.9966$ \\
max $\tau(\mbox{\ion{C}{1}}~\lambda10693)$\ldots & $0.23596$ & $0.02896$ & $0.00318$ & $0.00082$ & $0.00032$ & $0.00017$ \\
max $\tau(\mbox{\ion{C}{2}}~\lambda6580)$\dotfill & $10.07514$ & $1.99005$ & $0.83938$ & $0.30879$ & $0.05106$ & $0.01895$ \\
\sidehead{Outer part ($16000 < v \leq 19000{\rm\,km\,s^{-1}}$)}
$M_{\rm tot}~(M_\odot)$\dotfill & $0.0316$ & $0.0316$ & $0.0316$ & $0.0316$ & $0.0316$ & $0.0316$ \\
$M_{\rm C}~(M_\odot)$\dotfill & $0.0033$ & $0.0033$ & $0.0033$ & $0.0033$ & $0.0033$ & $0.0033$ \\
$M_{\rm C\, I}/M_{\rm C}$\dotfill & $0.001179$ & $0.000062$ & $0.000009$ & $0.000004$ & $0.000003$ & $0.000004$ \\
$M_{\rm C\, II}/M_{\rm C}$\dotfill & $0.9988$ & $0.9999$ & $0.9997$ & $0.9996$ & $0.9999$ & $0.9999$ \\
max $\tau(\mbox{\ion{C}{1}}~\lambda10693)$\ldots & $0.22876$ & $0.01900$ & $0.00216$ & $0.00056$ & $0.00018$ & $0.00009$ \\
max $\tau(\mbox{\ion{C}{2}}~\lambda6580)$\dotfill & $0.31105$ & $0.10694$ & $0.04343$ & $0.01573$ & $0.00264$ & $0.00101$ \\
\sidehead{Outskirts ($v > 19000{\rm\,km\,s^{-1}}$)}
$M_{\rm tot}~(M_\odot)$\dotfill & $0.0095$ & $0.0095$ & $0.0095$ & $0.0095$ & $0.0095$ & $0.0095$ \\
$M_{\rm C}~(M_\odot)$\dotfill & $0.0093$ & $0.0093$ & $0.0093$ & $0.0093$ & $0.0093$ & $0.0093$ \\
$M_{\rm C\, I}/M_{\rm C}$\dotfill & $0.004206$ & $0.000176$ & $0.000012$ & $0.000004$ & $0.000003$ & $0.000004$ \\
$M_{\rm C\, II}/M_{\rm C}$\dotfill & $0.9958$ & $0.9998$ & $1.0000$ & $1.0000$ & $1.0000$ & $1.0000$ \\
max $\tau(\mbox{\ion{C}{1}}~\lambda10693)$\ldots & $2.56294$ & $0.12246$ & $0.01079$ & $0.00248$ & $0.00073$ & $0.00037$ \\
max $\tau(\mbox{\ion{C}{2}}~\lambda6580)$\dotfill & $0.00463$ & $0.01429$ & $0.01587$ & $0.00880$ & $0.00243$ & $0.00088$ \\
\enddata
\tablenotetext{a}{All quantities for carbon are for a fiducial model where $(\xci , \xcm)=(0.002,0.01)$.}
\tablenotetext{b}{Relative to time of explosion, with maximum light at 19.1\,d.}
\end{deluxetable*}
\subsection{Where in the Ejecta is the Carbon Feature Formed?}
\label{subsec:physics_where}

To trace where a given feature is formed, we inspect the ejecta's relevant line opacities. In TARDIS, these are calculated in the Sobolev approximation, in which interaction is considered only when a photon packet is Doppler shifted into the line (as appropriate for SN~Ia ejecta with their large velocity gradients [\citealt{Sim2017_physics}]), and the code allows one to retrieve the Sobolev optical depth (i.e., the opacity integrated over the velocity width). 

In the top panel of Fig. \ref{Fig:physics} we show the Sobolev opacities of the \ion{C}{2} $\lambda$6580 line for our fiducial model.  One sees that the opacity peaks in a relatively small velocity range, between just above the pseudo-photosphere and $\sim\!14000{\rm\,km\,s^{-1}}$ (consistent with the findings of \citealt{Parrent2012_carbon}).

\subsection{Why does the C feature usually disappear near maximum light?}
\label{subsec:physics_disappear}

The line opacity depends on the column density of the relevant ions and will thus decrease with time as $t^{-2}$ due to the expansion of the ejecta if the level distributions do not change. Of course, these do also change and furthermore the amount of material involved in generating the line increases as the pseudo-photosphere recedes.

In the bottom panel of Fig.~\ref{Fig:physics}, we show how the fraction of ions at the relevant level for forming the carbon feature depends on the local density and temperature (under the NLTE approximations adopted here), with the  temperature-density profiles of SN~2011fe at different epochs overdrawn. One sees that for typical conditions, the hotter and denser regions are favored for forming the carbon feature; while in the hottest layers a significant fraction of carbon might be more than singly ionized (see Table~\ref{tb:quantities}), this is compensated by increased population of the relatively high excitation state from which the carbon feature arises.

One sees that near maximum, the inner regions of the ejecta are still sufficiently hot and dense that a significant fraction of carbon is in the relevant excitation state for the formation of the feature, but nevertheless the opacities have become quite small (see top panel of Fig. \ref{Fig:physics} and Table \ref{tb:quantities}). This indicates that the primary factor why the carbon feature weakens is the dilution of the column density due to the expansion of the ejecta. Nevertheless, at $t_{\rm{exp}}=16.1$ and 19.1\,d, the peak opacity is still of the order of 0.2, indicating that even though no carbon trough is formed, carbon still contributes to the flattening of the main silicon feature.

An important clue is provided by the fact that a carbon feature is almost never observed near maximum  in normal SNe~Ia, even though the density and temperature near the photosphere are typically ideal for forming the carbon signature. This implies that the inner regions of the ejecta must generally be carbon poor, with $\xc < 5\times 10^{-3}$ inferred from Fig.~\ref{Fig:plateaus} (where $\xci=5\times 10^{-3}$ does lead to the formation of a feature at $t_{\rm{exp}}=16.1$ and $19.1\,$d).

In this respect, an apparent exception to the rule may be informative: SN~2002fk did exhibit a carbon feature even one week past maximum \citep{Cartier2014_carbon}. This SN~Ia was relatively normal in its lightcurve, but stood out in that it exhibited a small expansion velocity ($v \lesssim 10000{\rm\,km\,s^{-1}}$, as measured near and post maximum from the \ion{Si}{2} $\lambda$6355 line). Using near-maximum velocities for a crude estimate, SN~2002fk's ejecta was slower by a factor of $\sim 0.94$ and thus, if conditions were otherwise similar to those of SN~2011fe, the column density in the former would be increased, and the time at which the carbon feature would disappear lengthened, by the inverse of that factor squared. This correction alone cannot account for carbon being observed at such late epochs in SN~2002fk, which suggests either even lower pre-maximum velocities or different conditions relatively to SN~2011fe.



\subsection{How much carbon can be hidden in normal SNe~Ia that do not exhibit a C feature?}
\label{subsec:physics_howmuch}

For objects similar to SN~2011fe, Fig. \ref{Fig:plateaus} indicates that towards maximum light, for inner carbon mass fractions $\xci\lesssim 10^{-3}$ there is little impact on the spectra, while for $10^{-3} \lesssim \xci\lesssim 2\ \times 10^{-3}$ the red wing of the main silicon feature is flattened, and for $\xci\gtrsim 5\ \times 10^{-3}$ there should be a detectable carbon feature. As expected, these ranges correspond to a transition from negligible to substantial optical depth ($\tau\lesssim 0.09$, $0.09 \lesssim\tau\lesssim 0.18$, and $\tau\gtrsim0.46$, resp.).

\subsection{Overluminous versus ``super-Chandrasekhar mass'' events}
\label{subsec:hot}

For the typical conditions in SN Ia ejecta, one generally expects that hotter and denser layers are more favorable for the formation of the carbon feature (see Fig. \ref{Fig:physics}, bottom panel). Given this, what can one learn for other sub-types of SN Ia?  The carbon feature is rarely observed in overluminous (91T-like) events \citep{Parrent2011_carbon}. These are thought to have somewhat higher temperatures than normal SN Ia like SN 2011fe, but otherwise similar density and velocity structures, which suggests the carbon feature should be, if anything, easier to detect: even if carbon were to become doubly ionized in the innermost layers, the conditions would still be favorable for forming the carbon feature in the layers above, which are generally more carbon rich. The lack of observed carbon signatures would then suggest that in these events the amount of unburned material is truly lower than in their normal counterparts.

On the other hand, for super-Chandrasekhar mass candidates, which are also associated with relatively hot ejecta, the carbon feature is one of the defining features. This would be unexpected if these objects were part of the same family: since they arise from even more energetic explosion than the 91T-like SN Ia, one would naively expect them to contain even less unburned material.  It is less clear whether the presence of carbon indicates a much larger abundance of unburned material, because the velocities inferred from the carbon feature in super-Chandrasekhar mass candidates are relatively low ($7500 \lesssim v \lesssim 10000{\rm\,km\,s^{-1}}$; \citealt{Taubenberger2017_subtypes}), and thus the carbon column density will be diluted relatively slowly (as discussed for SN~2002fk in Sect.~\ref{subsec:physics_disappear}).

Independent of the precise reasons, the above suggests a distinct explosion mechanism. It might be particularly fruitful to study overluminous and super-Chandrasekhar events together with possible  ``transitional'' events like SN~iPTF13asv (SN~2013cv, \citealt{Cao2016_subtype}). 

\section{Neutral Carbon}
\label{sec:neutral}

So far, we have only considered singly-ionized carbon, and in particular its $\lambda6580$ transition.  However, \citet{Hsiao2013_nir-carbon} presented evidence also for the presence of neutral carbon in SN~2011fe, from a flux deficit in the NIR attributed to \ion{C}{1} $\lambda10693$ (which has the strongest oscillator strength in the near-infrared region of the spectra; \citealt{Marion2006_carbon}), a flux deficit that seems to become more relevant at epochs near maximum (see their Fig. 4), and appears to move to lower velocity.

This detection is puzzling, since in our models carbon is mostly singly ionized throughout the ejecta (consistent with early work; e.g., \citealt{Tanaka2008_carbon}).  Indeed, we find that neutral carbon does not contribute significantly to our synthetic spectra, except possibly at very early times, $t_{\rm{exp}}=3.7$ and 5.9~d, where the optical depth in the \ion{C}{1} $\lambda10693$ feature reaches $2.6$ and $0.12$, respectively (see Table \ref{tb:quantities}).  But this happens only in the outermost layers, above $19000\ \rm{km\ s^{-1}}$, which, with  $\xc=0.97$, consist of almost pure carbon\footnote{M14 note that a C/O abundance ratio closer to 50/50 would worsen the fit to the oxygen features.}. The line opacity decreases drastically at later epochs; e.g., $\tau < 0.02$ for $t_{\rm{exp}}\geq 9\,$d and, consistently, we do not see clear signatures of \ion{C}{1} $\lambda10693$ in these simulated spectra.

We note that the fits are not particularly good in the NIR region, e.g., not reproducing the neighbouring Mg feature near $1.05\,\mu$m (see Fig.~\ref{Fig:scan}). This is possibly because the approximation of a sharp photosphere is less accurate to simulate the NIR region, implying that long-wavelength radiation from below the assumed photosphere could also interact with the ejecta above. Importantly, in our models, we find neutral carbon to have higher opacities in the cooler layers, which are located above the photosphere; Therefore, the approximation of a sharp photosphere is unlikely to affect the opacities we derive for neutral carbon. Also, we note that changing the abundance of Mg is unlikely to change the fact that carbon is predominantly ionized, rather than neutral, in our simulations. To verify this, we ran an additional model at  $t_{\rm{exp}}=9.0$\, d where the Mg abundance is increased by a factor of 10 in the carbon rich region; in this model, the opacity of the \ion{C}{1} $\lambda10693$ line remained small.

The above raises the question of whether neutral carbon is relevant or not in SN~2011fe and, more generally, in normal SNe~Ia. Could the identification by \citet{Hsiao2013_nir-carbon} be wrong? Or is the fraction of neutral carbon underestimated in TARDIS, perhaps because it does not fully take into account all NLTE effects?

In support of their identification, \citet{Hsiao2013_nir-carbon} use a physical argument first advanced by \citet{Marion2006_carbon}, that the fractions of both neutral carbon and neutral oxygen can be quite high because high opacities by Fe line transitions in the ultraviolet deplete the photons that can ionize these elements.  Here, \citet{Marion2006_carbon} noted that carbon and oxygen have similar ionization and excitation levels, and that the oscillator strength of the \ion{C}{1} $\lambda10693$ transition is about 50 times larger than that of the \ion{O}{1} $\lambda7773$ one.  They thus suggest that the detection of the \ion{O}{1} feature in SNe~Ia implies that, if any carbon is present in the same layers as oxygen, it would be partially neutral and produce a strong feature.

Since \citet{Marion2006_carbon} found no clear evidence of carbon in the objects they studied, they concluded that those events had low carbon-to-oxygen fractions.  But using the same argument that neutral carbon might be expected, \citet{Hsiao2013_nir-carbon}, \citet{Marion2015_nir-carbon} and \citet{Hsiao2015_iPTF13ehb} used line identification codes to infer that neutral carbon was present in the ejecta of SN~2011fe, SN~2014J and iPTF~13ebh, respectively, because it improved spectral fits in the NIR region.

As noted, our simulations do not show significant neutral carbon, but they do show an appreciable fraction of neutral oxygen. For instance, at $t_{\rm{exp}}=5.9$\,d, in the outer layers with $v\gtrsim 19000{\rm\,km\,s^{-1}}$ which have the most favorable conditions, about 10\% of oxygen but only 0.02\% of carbon is predicted to be neutral. Such differences are expected in thermal equilibrium despite carbon and oxygen having similar ionization potentials ($11.2$ and $13.6{\rm\,eV}$ respectively). For instance, for a typical temperature in the outermost layers of $T=7000{\rm\,K}$, the Saha equation estimates a neutral oxygen fraction about 100 times larger than the neutral carbon fraction.

To determine whether the identifications are correct, it may help to compare with sub-luminous SNe Ia, for which, given their cooler ejecta, the fraction of neutral carbon should be higher than in normal SNe~Ia.  Indeed, the subluminous SN~1999by was one of the first SNe~Ia for which neutral carbon was identified, by \citet{Hoflich2002_nir}, using a self-consistent radiation code that included a detailed NLTE treatment.

Interestingly, \citet{Hsiao2015_iPTF13ehb} detected neutral carbon in iPTF~13ebh, which is considered a transition object: it is on the fainter/cooler end of the \citet{Phillips1993_relation} relation, but its NIR maxima peaked before the $B$-band maximum and it did not exhibit a strong Ti trough.  The \ion{C}{1} $\lambda10693$ feature is strong in the earliest spectra (i.e. prior to $-10$\, d with respect to the $B$-band maximum), but quickly fades away, showing little velocity evolution. This behaviour is closer to what we would expect from cooler SN~2011fe models, where the temperature across the ejecta is artificially lowered so that the spectra resemble that of fainter SNe~Ia (as described in Paper I).  

Given the above, it is worth considering whether something else than \ion{C}{1} could cause the observed flux deficit near 1.03$\rm{\mu m}$ found in SN~2011fe and SN~2014J.  One alternative might be high velocity Mg.  This was discussed by \citet{Marion2015_nir-carbon} but discarded because of the lack of other high-velocity features at the same epoch (although the authors note that high-velocity Si might be present, but they cannot be sure because of possible blending with \ion{Na}{1}).

Alternatively, the flux deficit in the NIR might be due to helium. Studies of the normal SN~1994D by \citet{Meikle1996_He} and \citet{Mazzali1998_He} already hinted that \ion{He}{1} (and/or \ion{Mg}{2}) could impact the NIR spectra under NLTE conditions, although it was noted that the absence of this feature in post-maximum spectra set quite stringent constraints.  More recently, \citet{Boyle2017_He} investigated the possible signatures of \ion{He}{1} in NIR spectra in the context of double detonation models. Their synthetic spectra exhibited a stronger (deeper) He feature than what is seen in SN~2011fe (compare Fig.~4 of \citealt{Marion2015_nir-carbon} and Fig.~11 of \citealt{Boyle2017_He}), but perhaps this would be different for other progenitor models. Unfortunately, we cannot test this directly in our models, since for helium, with its unusual excitation levels, a more sophisticated NLTE treatment is necessary \citep{Boyle2017_He}.

From the discussion above, we conclude that in cooler SNe~Ia the \ion{C}{1} $\lambda10693$ signature has likely been detected; This is also corroborated by the identification of the other \ion{C}{1} features in SN~1999by and iPTF~13ebh. However, for normal SNe~Ia a more detailed investigation is needed, in which the relevant parameter space is explored with a code that can produce self-consistent spectra and adequately take into account NLTE effects.

\section{Model uncertainties}
\label{sec:uncertainties}

\begin{figure}
\epsscale{1.215}
\plotone{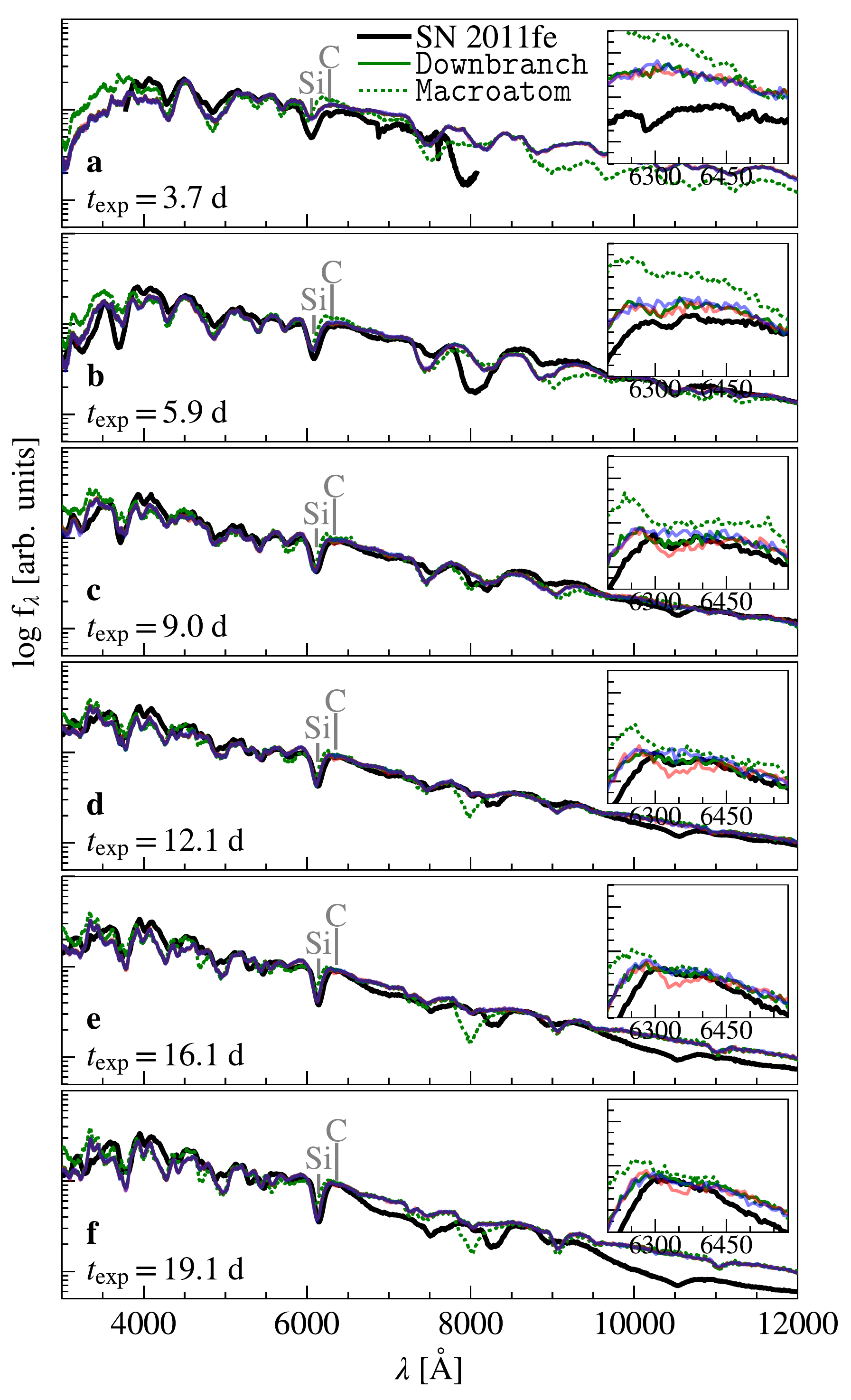}
\caption{Comparison of spectra calculated using different line treatments (green), with \texttt{downbranch} shown as full and \texttt{macroatom} as dotted lines. Epochs are as in other figures, $t_{\rm{exp}}=$ 3.7, 5.9, 9, 12.1, 16.1 and 19.1\,d (top to bottom).  For all, we adopted abundances as in our fiducial model,  $\xci=0.002$ and $\xcm=0.01$. The blue and red lines shows how the spectra would change if the boundary between the inner and middle regions at $v = 13500{\rm\,km\,s^{-1}}$ were moved by plus or minus $1500{\rm\,km\,s^{-1}}$, respectively. One sees that while the choice of line interaction treatment influences the flux levels, it has little impact on the strength and shape of the carbon feature and that our results are sensitive to position of the boundary between the two regions to a level of $1500{\rm\,km\,s^{-1}}$. Moving this boundary by $1000{\rm\,km\,s^{-1}}$ does not lead to appreciable changes in synthetic spectra (not shown).}
\label{Fig:best}
\end{figure}

Our simulated spectra have systematic uncertainties related to the approximations used in TARDIS. One is related to the line treatment. For the simulations shown here, we adopted the \texttt{downbranch} treatment, because this is the default approximation used in other Monte Carlo codes in the literature, thus allowing better comparisons.  In principle, however, the ``\texttt{macroatom}'' treatment is more physical, as it allows for de-excitation cascades and multiple-photon excitation. As already shown in \S 4 and Fig.~4 of paper I, the choice of line treatment does not have a large impact on spectra formed at the temperatures found in the ejecta of SN~2011fe. We find that also for this study the results do not depend on the treatment adopted.  To illustrate this, we show in Fig.~\ref{Fig:best} the spectra of our fiducial model for SN~2011fe with both line treatments. One sees that the main difference is in the flux level at the redder wavelengths\footnote{The different continuum shape means that with {\tt macroatom} the fits to the spectra are no longer correct, i.e., for a proper analysis the tomography would have to be redone.}; the shape of most spectral features are very similar.

Fig.~\ref{Fig:best} also illustrates the effect of changing the placement of the boundary between the inner and middle regions from the $v = 13500{\rm\,km\,s^{-1}}$ used in our analysis. This placement was somewhat arbitrarily taken from our ``depth scan'' (see Sect.~\ref{subsec:scan}). When we move this boundary to $v = 15000{\rm\,km\,s^{-1}}$ in our default model, where $\xci=0.002$ and $\xcm=0.01$, the carbon feature weakens earlier and is shallower than observed at $t_{\rm{exp}}=$ 9 and 12.1\,d. Conversely, when we place it at $v = 12000{\rm\,km\,s^{-1}}$, the carbon feature becomes stronger than observed for $t_{\rm{exp}}\geq$ 9\,d.  Both changes are as expected, and show that in interpreting our results, one has to keep in mind that our abundances are averages over fairly large regions.

Another possible source of uncertainty in our synthetic spectra arises from the incomplete treatment of NLTE effects. It is known that deviations from LTE can be important: for instance, \citet{Thomas2007_carbon} studied the CO-rich zone of their W7 model and found significant departures when computing the population of atomic levels of \ion{C}{2} that are relevant for producing lines in the optical. While \citeauthor{Thomas2007_carbon} do not mention the explicit velocities and epoch used to compute their departure coefficients, the CO-rich zone of W7 is at velocities $v\gtrsim 14000{\rm\,km\,s^{-1}}$ in the W7 model (and $v\gtrsim 19000{\rm\,km\,s^{-1}}$ in our simulations), where the densities and temperatures are the lowest and thus more likely to deviate from LTE.

Despite the CO-rich zone being composed of nearly $100\%$ carbon in our calculations, it still does not contribute significantly to the carbon feature.  This is consistent with the observations: for instance, using Fig.~4 of \citet{Parrent2012_carbon}, one can see that Doppler velocities of the carbon trough do not exceed $15000{\rm\,km\,s^{-1}}$.  Thus, NLTE effects are unlikely to greatly strengthen the carbon feature. We cannot exclude, however, the possibility that they greatly weaken it, in which case all our carbon mass fractions would be underestimates. Fortunately, for the deeper layers in which the feature is more likely to form, NLTE effects are likely to be less important.

Finally, as discussed in \S \ref{sec:neutral}, one specific NLTE effect is that line blanketing by Fe depletes the pool of UV photons that could otherwise ionize C and O, possibly causing our calculations to underestimate the amount of neutral carbon everywhere in the ejecta. TARDIS only uses a simple approximation to treat this problem, which is implemented (following \citealt{Mazzali1993_code}) as a correction factor $\delta$ when estimating the balance of ions under the \texttt{nebular} ionization mode. We tested the influence of this approximation by comparing our fiducial spectra to models that enforce $\delta=1$ (i.e. no correction). The spectra did not differ in any significant way and thus this does help to determine neither whether depletion of UV photons is accounted for properly, nor whether the fraction of neutral carbon is estimated correctly in TARDIS.  

Despite these uncertainties, we stress that compared to work that relies on line identification routines (e.g. \citealt{Parrent2011_carbon,Parrent2012_carbon,Hsiao2013_nir-carbon,Marion2015_nir-carbon}), our approach has the advantage of computing temperatures and ionization balances in a more consistent way. Compared to full tomographic analysis (e.g. \citealt{Stehle2005_tomography,Mazzali2008_tomography,Tanaka2011_tomography,Sasdelli2014_tomography,Ashall2016_tomography}), our focus on one particular element helps to derive not just approximate values but also get a sense of the range of acceptable values. Nevertheless, in all these approaches, for a more complete assessment of their results, it will be important to compare with detailed NTLE treatments.

\section{Conclusions and Ramifications}
\label{sec:ramifications}

In this work we have used a previously published tomographic analysis \citep{Mazzali2014_tomography} to investigate in detail the distribution of carbon at depth in the ejecta of SN~2011fe.

Based on the local temperatures and densities in the ejecta of SN~2011fe, we find that the carbon feature should be strong if significant amounts of carbon are present near the photosphere for epochs up to maximum. We thus interpret the lack of observed events with a clear feature near maximum as evidence for the lack of carbon deep in the ejecta.

We also find that for the relevant conditions, carbon is predominantly singly ionized and unlikely to produce the near-infared \ion{C}{1} $\lambda10693$ line; This agrees with the findings of \citet{Tanaka2008_carbon} and leaves as a bit of a mystery the flux deficit near $1.03{\rm\,\mu m}$ observed in some normal SN~Ia. It remains an open question whether it is due to neutral carbon (if full NLTE effects are relevant), neutral He \citep{Boyle2017_He} or high velocity Mg \citep{Marion2015_nir-carbon}.

We compare our results with predictions for various SN Ia explosion scenarios and find that the violent merger model of \citet{Pakmor2012_DD} predicts, on average, too much carbon mixed across the whole ejecta. In contrast, the spontaneous detonation of cold white dwarfs with no He shell, which might be seen as a first-order approximation to, e.g., the dynamically-driven double-detonation model of \citet{Shen2018_ddet}, produce too little carbon at the velocities where we detect it. The discrepancies for these models are so large (factor of $\sim\!10$) that we believe these are excluded even taking into account the approximations used in our approach.

\textbf{The delayed detonation model of \citet{Seitenzahl2013_DDT} and the gravitationally confined detonation model of \citet{Seitenzahl2016_gcd}} are roughly consistent with our results, but might well be excluded by determining the distribution of unburned material in sub- and over-luminous events, or by investigating in detail the outermost layers of the ejecta. For instance, the double detonation models predict that the carbon abundance should peak at intermediate velocities, with the velocity of maximum abundance likely depending on mass of the exploding white dwarf (and thus peak luminosity; \citealt{Fink2010_helium})\textbf{, whereas in DDT models, carbon abundance is predicted to be higher for fainter SNe Ia (e.g. \citealt{Wheeler1998_nir, Hoflich2002_nir}).} 

\acknowledgments
We thank Stephen Ro for insightful discussion regarding the physics of exploding WDs. This work made use of the Heidelberg Supernova Model Archive (HESMA), https://hesma.h-its.org.
W. E. Kerzendorf was supported by an ESO Fellowship and the Excellence Cluster Universe, Technische Universit{\"a}t
M{\"u}nchen, Boltzmannstrasse 2, D-85748 Garching, Germany. SAS acknowledges support from STFC via grant ST/P000312/1. 

\software{Astropy \citep{Astropy2013}, TARDIS \citep{Kerzendorf2014_TARDIS, Kerzendorf2018_TARDIS}}.

\bibliographystyle{./apj}

\begin{thebibliography}{}
\expandafter\ifx\csname natexlab\endcsname\relax\def\natexlab#1{#1}\fi

\bibitem[{{Ashall} {et~al.}(2016){Ashall}, {Mazzali}, {Pian}, \&
  {James}}]{Ashall2016_tomography}
{Ashall}, C., {Mazzali}, P.~A., {Pian}, E., \& {James}, P.~A. 2016, \mnras,
  463, 1891

\bibitem[{{Astropy Collaboration} {et~al.}(2013){Astropy Collaboration},
  {Robitaille}, {Tollerud}, {Greenfield}, {Droettboom}, {Bray}, {Aldcroft},
  {Davis}, {Ginsburg}, {Price-Whelan}, {Kerzendorf}, {Conley}, {Crighton},
  {Barbary}, {Muna}, {Ferguson}, {Grollier}, {Parikh}, {Nair}, {Unther},
  {Deil}, {Woillez}, {Conseil}, {Kramer}, {Turner}, {Singer}, {Fox}, {Weaver},
  {Zabalza}, {Edwards}, {Azalee Bostroem}, {Burke}, {Casey}, {Crawford},
  {Dencheva}, {Ely}, {Jenness}, {Labrie}, {Lim}, {Pierfederici}, {Pontzen},
  {Ptak}, {Refsdal}, {Servillat}, \& {Streicher}}]{Astropy2013}
{Astropy Collaboration}, {Robitaille}, T.~P., {Tollerud}, E.~J., {et~al.} 2013,
  \aap, 558, A33

\bibitem[{{Blondin} {et~al.}(2013){Blondin}, {Dessart}, {Hillier}, \&
  {Khokhlov}}]{Blondin2013_DDT}
{Blondin}, S., {Dessart}, L., {Hillier}, D.~J., \& {Khokhlov}, A.~M. 2013,
  \mnras, 429, 2127

\bibitem[{{Boyle} {et~al.}(2017){Boyle}, {Sim}, {Hachinger}, \&
  {Kerzendorf}}]{Boyle2017_He}
{Boyle}, A., {Sim}, S.~A., {Hachinger}, S., \& {Kerzendorf}, W. 2017, \aap,
  599, A46

\bibitem[{{Branch} {et~al.}(2002){Branch}, {Benetti}, {Kasen}, {Baron},
  {Jeffery}, {Hatano}, {Stathakis}, {Filippenko}, {Matheson}, {Pastorello},
  {Altavilla}, {Cappellaro}, {Rizzi}, {Turatto}, {Li}, {Leonard}, \&
  {Shields}}]{Branch2002_SYNOW}
{Branch}, D., {Benetti}, S., {Kasen}, D., {et~al.} 2002, \apj, 566, 1005

\bibitem[{{Branch} {et~al.}(2003){Branch}, {Garnavich}, {Matheson}, {Baron},
  {Thomas}, {Hatano}, {Challis}, {Jha}, \& {Kirshner}}]{Branch2003_carbon}
{Branch}, D., {Garnavich}, P., {Matheson}, T., {et~al.} 2003, \aj, 126, 1489

\bibitem[{{Branch} {et~al.}(2006){Branch}, {Dang}, {Hall}, {Ketchum},
  {Melakayil}, {Parrent}, {Troxel}, {Casebeer}, {Jeffery}, \&
  {Baron}}]{Branch2006_subclasses}
{Branch}, D., {Dang}, L.~C., {Hall}, N., {et~al.} 2006, \pasp, 118, 560

\bibitem[{{Bulla} {et~al.}(2016){Bulla}, {Sim}, {Pakmor}, {Kromer},
  {Taubenberger}, {R{\"o}pke}, {Hillebrandt}, \&
  {Seitenzahl}}]{Bulla2016_VMpol}
{Bulla}, M., {Sim}, S.~A., {Pakmor}, R., {et~al.} 2016, \mnras, 455, 1060

\bibitem[{{Cao} {et~al.}(2016){Cao}, {Johansson}, {Nugent}, {Goobar}, {Nordin},
  {Kulkarni}, {Cenko}, {Fox}, {Kasliwal}, {Fremling}, {Amanullah}, {Hsiao},
  {Perley}, {Bue}, {Masci}, {Lee}, \& {Chotard}}]{Cao2016_subtype}
{Cao}, Y., {Johansson}, J., {Nugent}, P.~E., {et~al.} 2016, \apj, 823, 147

\bibitem[{{Cartier} {et~al.}(2014){Cartier}, {Hamuy}, {Pignata}, {F{\"o}rster},
  {Zelaya}, {Folatelli}, {Phillips}, {Morrell}, {Krisciunas}, {Suntzeff},
  {Clocchiatti}, {Coppi}, {Contreras}, {Roth}, {Koviak}, {Maza},
  {Gonz{\'a}lez}, {Gonz{\'a}lez}, \& {Huerta}}]{Cartier2014_carbon}
{Cartier}, R., {Hamuy}, M., {Pignata}, G., {et~al.} 2014, \apj, 789, 89

\bibitem[{{Chomiuk} {et~al.}(2016){Chomiuk}, {Soderberg}, {Chevalier},
  {Bruzewski}, {Foley}, {Parrent}, {Strader}, {Badenes}, {Fransson}, {Kamble},
  {Margutti}, {Rupen}, \& {Simon}}]{Chomiuk2016_subtypes}
{Chomiuk}, L., {Soderberg}, A.~M., {Chevalier}, R.~A., {et~al.} 2016, \apj,
  821, 119

\bibitem[{{Ciaraldi-Schoolmann} {et~al.}(2013){Ciaraldi-Schoolmann},
  {Seitenzahl}, \& {R{\"o}pke}}]{Ciaraldi2013A&ADTD}
{Ciaraldi-Schoolmann}, F., {Seitenzahl}, I.~R., \& {R{\"o}pke}, F.~K. 2013,
  \aap, 559, A117

\bibitem[{{Fink} {et~al.}(2018){Fink}, {Kromer}, {Hillebrandt}, {Roepke},
  {Pakmor}, {Seitenzahl}, \& {Sim}}]{Fink2018_diffrot}
{Fink}, M., {Kromer}, M., {Hillebrandt}, W., {et~al.} 2018, ArXiv e-prints,
  arXiv:1807.10199

\bibitem[{{Fink} {et~al.}(2010){Fink}, {R{\"o}pke}, {Hillebrandt},
  {Seitenzahl}, {Sim}, \& {Kromer}}]{Fink2010_helium}
{Fink}, M., {R{\"o}pke}, F.~K., {Hillebrandt}, W., {et~al.} 2010, \aap, 514,
  A53

\bibitem[{{Fink} {et~al.}(2014){Fink}, {Kromer}, {Seitenzahl},
  {Ciaraldi-Schoolmann}, {R{\"o}pke}, {Sim}, {Pakmor}, {Ruiter}, \&
  {Hillebrandt}}]{Fink2014_def}
{Fink}, M., {Kromer}, M., {Seitenzahl}, I.~R., {et~al.} 2014, \mnras, 438, 1762

\bibitem[{{Folatelli} {et~al.}(2012){Folatelli}, {Phillips}, {Morrell},
  {Tanaka}, {Maeda}, {Nomoto}, {Stritzinger}, {Burns}, {Hamuy}, {Mazzali},
  {Boldt}, {Campillay}, {Contreras}, {Gonz{\'a}lez}, {Roth}, {Salgado},
  {Freedman}, {Madore}, {Persson}, \& {Suntzeff}}]{Folatelli2012_carbon}
{Folatelli}, G., {Phillips}, M.~M., {Morrell}, N., {et~al.} 2012, \apj, 745, 74

\bibitem[{{Garavini} {et~al.}(2005){Garavini}, {Aldering}, {Amadon},
  {Amanullah}, {Astier}, {Balland}, {Blanc}, {Conley}, {Dahl{\'e}n}, {Deustua},
  {Ellis}, {Fabbro}, {Fadeyev}, {Fan}, {Folatelli}, {Frye}, {Gates}, {Gibbons},
  {Goldhaber}, {Goldman}, {Goobar}, {Groom}, {Haissinski}, {Hardin}, {Hook},
  {Howell}, {Kent}, {Kim}, {Knop}, {Kowalski}, {Kuznetsova}, {Lee}, {Lidman},
  {Mendez}, {Miller}, {Moniez}, {Mouchet}, {Mour{\~a}o}, {Newberg}, {Nobili},
  {Nugent}, {Pain}, {Perdereau}, {Perlmutter}, {Quimby}, {Regnault}, {Rich},
  {Richards}, {Ruiz-Lapuente}, {Schaefer}, {Schahmaneche}, {Smith},
  {Spadafora}, {Stanishev}, {Thomas}, {Walton}, {Wang}, {Wood-Vasey}, \&
  {Supernova Cosmology Project}}]{Garavini2005_carbon}
{Garavini}, G., {Aldering}, G., {Amadon}, A., {et~al.} 2005, \aj, 130, 2278

\bibitem[{{Garavini} {et~al.}(2007){Garavini}, {Folatelli}, {Nobili},
  {Aldering}, {Amanullah}, {Antilogus}, {Astier}, {Blanc}, {Bronder}, {Burns},
  {Conley}, {Deustua}, {Doi}, {Fabbro}, {Fadeyev}, {Gibbons}, {Goldhaber},
  {Goobar}, {Groom}, {Hook}, {Howell}, {Kashikawa}, {Kim}, {Kowalski},
  {Kuznetsova}, {Lee}, {Lidman}, {Mendez}, {Morokuma}, {Motohara}, {Nugent},
  {Pain}, {Perlmutter}, {Quimby}, {Raux}, {Regnault}, {Ruiz-Lapuente},
  {Sainton}, {Schahmaneche}, {Smith}, {Spadafora}, {Stanishev}, {Thomas},
  {Walton}, {Wang}, {Wood-Vasey}, \& {Yasuda}}]{Garavini2007_pEW}
{Garavini}, G., {Folatelli}, G., {Nobili}, S., {et~al.} 2007, \aap, 470, 411

\bibitem[{{Hachinger} {et~al.}(2012){Hachinger}, {Mazzali}, {Taubenberger},
  {Fink}, {Pakmor}, {Hillebrandt}, \& {Seitenzahl}}]{Hachinger2012_tomography}
{Hachinger}, S., {Mazzali}, P.~A., {Taubenberger}, S., {et~al.} 2012, \mnras,
  427, 2057

\bibitem[{{Heringer} {et~al.}(2017){Heringer}, {van Kerkwijk}, {Sim}, \&
  {Kerzendorf}}]{Heringer2017_sequence}
{Heringer}, E., {van Kerkwijk}, M.~H., {Sim}, S.~A., \& {Kerzendorf}, W.~E.
  2017, \apj, 846, 15

\bibitem[{{Hoeflich}(1990)}]{Hoeflich1990_brightening}
{Hoeflich}, P. 1990, \aap, 229, 191

\bibitem[{{Hoeflich} {et~al.}(1995){Hoeflich}, {Khokhlov}, \&
  {Wheeler}}]{Hoflich1995_radiative}
{Hoeflich}, P., {Khokhlov}, A.~M., \& {Wheeler}, J.~C. 1995, \apj, 444, 831

\bibitem[{{H{\"o}flich} {et~al.}(2002){H{\"o}flich}, {Gerardy}, {Fesen}, \&
  {Sakai}}]{Hoflich2002_nir}
{H{\"o}flich}, P., {Gerardy}, C.~L., {Fesen}, R.~A., \& {Sakai}, S. 2002, \apj,
  568, 791

\bibitem[{{Hsiao} {et~al.}(2013){Hsiao}, {Marion}, {Phillips}, {Burns},
  {Winge}, {Morrell}, {Contreras}, {Freedman}, {Kromer}, {Gall}, {Gerardy},
  {H{\"o}flich}, {Im}, {Jeon}, {Kirshner}, {Nugent}, {Persson}, {Pignata},
  {Roth}, {Stanishev}, {Stritzinger}, \& {Suntzeff}}]{Hsiao2013_nir-carbon}
{Hsiao}, E.~Y., {Marion}, G.~H., {Phillips}, M.~M., {et~al.} 2013, \apj, 766,
  72

\bibitem[{{Hsiao} {et~al.}(2015){Hsiao}, {Burns}, {Contreras}, {H{\"o}flich},
  {Sand}, {Marion}, {Phillips}, {Stritzinger}, {Gonz{\'a}lez-Gait{\'a}n},
  {Mason}, {Folatelli}, {Parent}, {Gall}, {Amanullah}, {Anupama}, {Arcavi},
  {Banerjee}, {Beletsky}, {Blanc}, {Bloom}, {Brown}, {Campillay}, {Cao}, {De
  Cia}, {Diamond}, {Freedman}, {Gonzalez}, {Goobar}, {Holmbo}, {Howell},
  {Johansson}, {Kasliwal}, {Kirshner}, {Krisciunas}, {Kulkarni}, {Maguire},
  {Milne}, {Morrell}, {Nugent}, {Ofek}, {Osip}, {Palunas}, {Perley}, {Persson},
  {Piro}, {Rabus}, {Roth}, {Schiefelbein}, {Srivastav}, {Sullivan}, {Suntzeff},
  {Surace}, {Wo{\'z}niak}, \& {Yaron}}]{Hsiao2015_iPTF13ehb}
{Hsiao}, E.~Y., {Burns}, C.~R., {Contreras}, C., {et~al.} 2015, \aap, 578, A9

\bibitem[{{Iwamoto} {et~al.}(1999){Iwamoto}, {Brachwitz}, {Nomoto},
  {Kishimoto}, {Umeda}, {Hix}, \& {Thielemann}}]{Iwamoto1999_WDD}
{Iwamoto}, K., {Brachwitz}, F., {Nomoto}, K., {et~al.} 1999, \apjs, 125, 439

\bibitem[{Kerzendorf {et~al.}(2018)Kerzendorf, Nöbauer, Sim, Lietzau,
  Jančauskas, Vogl, Mishin, Tsamis, Boyle, Gupta, Desai, Klauser, Beaujean,
  Suban-Loewen, Heringer, Shingles, Barna, Gautam, Patel, Barbosa, Varanasi,
  Reinecke, Bylund, Bentil, Rajagopalan, Jain, Singh, Talegaonkar, Sofiatti,
  Patel, Yap, Wahi, \& Gupta}]{Kerzendorf2018_TARDIS}
Kerzendorf, W., Nöbauer, U., Sim, S., {et~al.} 2018, tardis-sn/tardis: TARDIS
  v2.0.2 release, doi:10.5281/zenodo.1292315

\bibitem[{{Kerzendorf} \& {Sim}(2014)}]{Kerzendorf2014_TARDIS}
{Kerzendorf}, W.~E., \& {Sim}, S.~A. 2014, \mnras, 440, 387

\bibitem[{{Khokhlov}(1991)}]{Khokhlov1991_DDT}
{Khokhlov}, A.~M. 1991, \aap, 245, 114

\bibitem[{Kramida {et~al.}(2018)Kramida, {Yu.~Ralchenko}, Reader, \& {and NIST
  ASD Team}}]{NIST_ASD}
Kramida, A., {Yu.~Ralchenko}, Reader, J., \& {and NIST ASD Team}. 2018, {NIST
  Atomic Spectra Database (ver. 5.5.2), [Online]. Available:
  {\tt{https://physics.nist.gov/asd}} [2018, February 27]. National Institute
  of Standards and Technology, Gaithersburg, MD.}

\bibitem[{{Kromer} {et~al.}(2017){Kromer}, {Ohlmann}, \&
  {R{\"o}pke}}]{Kromer2017_hesma}
{Kromer}, M., {Ohlmann}, S., \& {R{\"o}pke}, F.~K. 2017, \memsai, 88, 312

\bibitem[{{Kromer} {et~al.}(2010){Kromer}, {Sim}, {Fink}, {R{\"o}pke},
  {Seitenzahl}, \& {Hillebrandt}}]{Kromer2010_doubledet}
{Kromer}, M., {Sim}, S.~A., {Fink}, M., {et~al.} 2010, \apj, 719, 1067

\bibitem[{{Livio} \& {Mazzali}(2018)}]{Livio2018_review}
{Livio}, M., \& {Mazzali}, P. 2018, \physrep, 736, 1

\bibitem[{{Livne}(1990)}]{Livne1990_doubledet}
{Livne}, E. 1990, \apjl, 354, L53

\bibitem[{{Lucy}(1999)}]{Lucy1999_code}
{Lucy}, L.~B. 1999, \aap, 345, 211

\bibitem[{{Marion} {et~al.}(2006){Marion}, {H{\"o}flich}, {Wheeler},
  {Robinson}, {Gerardy}, \& {Vacca}}]{Marion2006_carbon}
{Marion}, G.~H., {H{\"o}flich}, P., {Wheeler}, J.~C., {et~al.} 2006, \apj, 645,
  1392

\bibitem[{{Marion} {et~al.}(2015){Marion}, {Sand}, {Hsiao}, {Banerjee},
  {Valenti}, {Stritzinger}, {Vink{\'o}}, {Joshi}, {Venkataraman}, {Ashok},
  {Amanullah}, {Binzel}, {Bochanski}, {Bryngelson}, {Burns}, {Drozdov},
  {Fieber-Beyer}, {Graham}, {Howell}, {Johansson}, {Kirshner}, {Milne},
  {Parrent}, {Silverman}, {Vervack}, \& {Wheeler}}]{Marion2015_nir-carbon}
{Marion}, G.~H., {Sand}, D.~J., {Hsiao}, E.~Y., {et~al.} 2015, \apj, 798, 39

\bibitem[{{Maund} {et~al.}(2013){Maund}, {Spyromilio}, {H{\"o}flich},
  {Wheeler}, {Baade}, {Clocchiatti}, {Patat}, {Reilly}, {Wang}, \&
  {Zelaya}}]{Maund2013_VMpol}
{Maund}, J.~R., {Spyromilio}, J., {H{\"o}flich}, P.~A., {et~al.} 2013, \mnras,
  433, L20

\bibitem[{{Mazzali} \& {Lucy}(1993)}]{Mazzali1993_code}
{Mazzali}, P.~A., \& {Lucy}, L.~B. 1993, \aap, 279, 447

\bibitem[{{Mazzali} \& {Lucy}(1998)}]{Mazzali1998_He}
---. 1998, \mnras, 295, 428

\bibitem[{{Mazzali} {et~al.}(2008){Mazzali}, {Sauer}, {Pastorello}, {Benetti},
  \& {Hillebrandt}}]{Mazzali2008_tomography}
{Mazzali}, P.~A., {Sauer}, D.~N., {Pastorello}, A., {Benetti}, S., \&
  {Hillebrandt}, W. 2008, \mnras, 386, 1897

\bibitem[{{Mazzali} {et~al.}(2014){Mazzali}, {Sullivan}, {Hachinger}, {Ellis},
  {Nugent}, {Howell}, {Gal-Yam}, {Maguire}, {Cooke}, {Thomas}, {Nomoto}, \&
  {Walker}}]{Mazzali2014_tomography}
{Mazzali}, P.~A., {Sullivan}, M., {Hachinger}, S., {et~al.} 2014, \mnras, 439,
  1959

\bibitem[{{Meikle} {et~al.}(1996){Meikle}, {Cumming}, {Geballe}, {Lewis},
  {Walton}, {Balcells}, {Cimatti}, {Croom}, {Dhillon}, {Economou}, {Jenkins},
  {Knapen}, {Meadows}, {Morris}, {Perez-Fournon}, {Shanks}, {Smith}, {Tanvir},
  {Veilleux}, {Vilchez}, {Wall}, \& {Lucey}}]{Meikle1996_He}
{Meikle}, W.~P.~S., {Cumming}, R.~J., {Geballe}, T.~R., {et~al.} 1996, \mnras,
  281, 263

\bibitem[{{Nomoto} {et~al.}(1984){Nomoto}, {Thielemann}, \&
  {Yokoi}}]{Nomoto1984_W7}
{Nomoto}, K., {Thielemann}, F.-K., \& {Yokoi}, K. 1984, \apj, 286, 644

\bibitem[{{Nugent} {et~al.}(2011){Nugent}, {Sullivan}, {Cenko}, {Thomas},
  {Kasen}, {Howell}, {Bersier}, {Bloom}, {Kulkarni}, {Kandrashoff},
  {Filippenko}, {Silverman}, {Marcy}, {Howard}, {Isaacson}, {Maguire},
  {Suzuki}, {Tarlton}, {Pan}, {Bildsten}, {Fulton}, {Parrent}, {Sand},
  {Podsiadlowski}, {Bianco}, {Dilday}, {Graham}, {Lyman}, {James}, {Kasliwal},
  {Law}, {Quimby}, {Hook}, {Walker}, {Mazzali}, {Pian}, {Ofek}, {Gal-Yam}, \&
  {Poznanski}}]{Nugent2011_WD}
{Nugent}, P.~E., {Sullivan}, M., {Cenko}, S.~B., {et~al.} 2011, \nat, 480, 344

\bibitem[{{Pakmor} {et~al.}(2012){Pakmor}, {Kromer}, {Taubenberger}, {Sim},
  {R{\"o}pke}, \& {Hillebrandt}}]{Pakmor2012_DD}
{Pakmor}, R., {Kromer}, M., {Taubenberger}, S., {et~al.} 2012, \apjl, 747, L10

\bibitem[{{Parrent} {et~al.}(2011){Parrent}, {Thomas}, {Fesen}, {Marion},
  {Challis}, {Garnavich}, {Milisavljevic}, {Vink{\`o}}, \&
  {Wheeler}}]{Parrent2011_carbon}
{Parrent}, J.~T., {Thomas}, R.~C., {Fesen}, R.~A., {et~al.} 2011, \apj, 732, 30

\bibitem[{{Parrent} {et~al.}(2012){Parrent}, {Howell}, {Friesen}, {Thomas},
  {Fesen}, {Milisavljevic}, {Bianco}, {Dilday}, {Nugent}, {Baron}, {Arcavi},
  {Ben-Ami}, {Bersier}, {Bildsten}, {Bloom}, {Cao}, {Cenko}, {Filippenko},
  {Gal-Yam}, {Kasliwal}, {Konidaris}, {Kulkarni}, {Law}, {Levitan}, {Maguire},
  {Mazzali}, {Ofek}, {Pan}, {Polishook}, {Poznanski}, {Quimby}, {Silverman},
  {Sternberg}, {Sullivan}, {Walker}, {Xu}, {Buton}, \&
  {Pereira}}]{Parrent2012_carbon}
{Parrent}, J.~T., {Howell}, D.~A., {Friesen}, B., {et~al.} 2012, \apjl, 752,
  L26

\bibitem[{{Pastorello} {et~al.}(2007){Pastorello}, {Taubenberger},
  {Elias-Rosa}, {Mazzali}, {Pignata}, {Cappellaro}, {Garavini}, {Nobili},
  {Anupama}, {Bayliss}, {Benetti}, {Bufano}, {Chakradhari}, {Kotak}, {Goobar},
  {Navasardyan}, {Patat}, {Sahu}, {Salvo}, {Schmidt}, {Stanishev}, {Turatto},
  \& {Hillebrandt}}]{Pastorello2005_05cf}
{Pastorello}, A., {Taubenberger}, S., {Elias-Rosa}, N., {et~al.} 2007, \mnras,
  376, 1301

\bibitem[{{Phillips}(1993)}]{Phillips1993_relation}
{Phillips}, M.~M. 1993, \apjl, 413, L105

\bibitem[{{Piro} \& {Nakar}(2013)}]{Piro_darkphase}
{Piro}, A.~L., \& {Nakar}, E. 2013, \apj, 769, 67

\bibitem[{{Plewa} {et~al.}(2004){Plewa}, {Calder}, \& {Lamb}}]{Plewa2004_GCD}
{Plewa}, T., {Calder}, A.~C., \& {Lamb}, D.~Q. 2004, \apjl, 612, L37

\bibitem[{{R{\"o}pke} {et~al.}(2012){R{\"o}pke}, {Kromer}, {Seitenzahl},
  {Pakmor}, {Sim}, {Taubenberger}, {Ciaraldi-Schoolmann}, {Hillebrandt},
  {Aldering}, {Antilogus}, {Baltay}, {Benitez-Herrera}, {Bongard}, {Buton},
  {Canto}, {Cellier-Holzem}, {Childress}, {Chotard}, {Copin}, {Fakhouri},
  {Fink}, {Fouchez}, {Gangler}, {Guy}, {Hachinger}, {Hsiao}, {Chen},
  {Kerschhaggl}, {Kowalski}, {Nugent}, {Paech}, {Pain}, {Pecontal}, {Pereira},
  {Perlmutter}, {Rabinowitz}, {Rigault}, {Runge}, {Saunders}, {Smadja},
  {Suzuki}, {Tao}, {Thomas}, {Tilquin}, \& {Wu}}]{Ropke2012_N100}
{R{\"o}pke}, F.~K., {Kromer}, M., {Seitenzahl}, I.~R., {et~al.} 2012, \apjl,
  750, L19

\bibitem[{{Sasdelli} {et~al.}(2014){Sasdelli}, {Mazzali}, {Pian}, {Nomoto},
  {Hachinger}, {Cappellaro}, \& {Benetti}}]{Sasdelli2014_tomography}
{Sasdelli}, M., {Mazzali}, P.~A., {Pian}, E., {et~al.} 2014, \mnras, 445, 711

\bibitem[{{Seitenzahl} {et~al.}(2013){Seitenzahl}, {Ciaraldi-Schoolmann},
  {R{\"o}pke}, {Fink}, {Hillebrandt}, {Kromer}, {Pakmor}, {Ruiter}, {Sim}, \&
  {Taubenberger}}]{Seitenzahl2013_DDT}
{Seitenzahl}, I.~R., {Ciaraldi-Schoolmann}, F., {R{\"o}pke}, F.~K., {et~al.}
  2013, \mnras, 429, 1156

\bibitem[{{Seitenzahl} {et~al.}(2016){Seitenzahl}, {Kromer}, {Ohlmann},
  {Ciaraldi-Schoolmann}, {Marquardt}, {Fink}, {Hillebrandt}, {Pakmor},
  {R{\"o}pke}, {Ruiter}, {Sim}, \& {Taubenberger}}]{Seitenzahl2016_gcd}
{Seitenzahl}, I.~R., {Kromer}, M., {Ohlmann}, S.~T., {et~al.} 2016, \aap, 592,
  A57

\bibitem[{{Shen} {et~al.}(2018){Shen}, {Kasen}, {Miles}, \&
  {Townsley}}]{Shen2018_ddet}
{Shen}, K.~J., {Kasen}, D., {Miles}, B.~J., \& {Townsley}, D.~M. 2018, \apj,
  854, 52

\bibitem[{{Shen} \& {Moore}(2014)}]{Shen2014_ddet}
{Shen}, K.~J., \& {Moore}, K. 2014, \apj, 797, 46

\bibitem[{{Shigeyama} {et~al.}(1992){Shigeyama}, {Nomoto}, {Yamaoka}, \&
  {Thielemann}}]{Shigeyama1992_subChandra}
{Shigeyama}, T., {Nomoto}, K., {Yamaoka}, H., \& {Thielemann}, F.-K. 1992,
  \apjl, 386, L13

\bibitem[{{Silverman} \& {Filippenko}(2012)}]{Silverman2012_BSNIP_IV}
{Silverman}, J.~M., \& {Filippenko}, A.~V. 2012, \mnras, 425, 1917

\bibitem[{{Sim}(2017)}]{Sim2017_physics}
{Sim}, S.~A. 2017, {Spectra of Supernovae During the Photospheric Phase}, ed.
  A.~W. {Alsabti} \& P.~{Murdin}, 769

\bibitem[{{Sim} {et~al.}(2012){Sim}, {Fink}, {Kromer}, {R{\"o}pke}, {Ruiter},
  \& {Hillebrandt}}]{Sim2012_doubledet}
{Sim}, S.~A., {Fink}, M., {Kromer}, M., {et~al.} 2012, \mnras, 420, 3003

\bibitem[{{Sim} {et~al.}(2010){Sim}, {R{\"o}pke}, {Hillebrandt}, {Kromer},
  {Pakmor}, {Fink}, {Ruiter}, \& {Seitenzahl}}]{Sim2010_subChandra}
{Sim}, S.~A., {R{\"o}pke}, F.~K., {Hillebrandt}, W., {et~al.} 2010, \apjl, 714,
  L52

\bibitem[{{Sim} {et~al.}(2013){Sim}, {Seitenzahl}, {Kromer},
  {Ciaraldi-Schoolmann}, {R{\"o}pke}, {Fink}, {Hillebrandt}, {Pakmor},
  {Ruiter}, \& {Taubenberger}}]{Sim2013_radtransfer}
{Sim}, S.~A., {Seitenzahl}, I.~R., {Kromer}, M., {et~al.} 2013, \mnras, 436,
  333

\bibitem[{{Stehle} {et~al.}(2005){Stehle}, {Mazzali}, {Benetti}, \&
  {Hillebrandt}}]{Stehle2005_tomography}
{Stehle}, M., {Mazzali}, P.~A., {Benetti}, S., \& {Hillebrandt}, W. 2005,
  \mnras, 360, 1231

\bibitem[{{Tanaka} {et~al.}(2011){Tanaka}, {Mazzali}, {Stanishev}, {Maurer},
  {Kerzendorf}, \& {Nomoto}}]{Tanaka2011_tomography}
{Tanaka}, M., {Mazzali}, P.~A., {Stanishev}, V., {et~al.} 2011, \mnras, 410,
  1725

\bibitem[{{Tanaka} {et~al.}(2008){Tanaka}, {Mazzali}, {Benetti}, {Nomoto},
  {Elias-Rosa}, {Kotak}, {Pignata}, {Stanishev}, \&
  {Hachinger}}]{Tanaka2008_carbon}
{Tanaka}, M., {Mazzali}, P.~A., {Benetti}, S., {et~al.} 2008, \apj, 677, 448

\bibitem[{{Taubenberger}(2017)}]{Taubenberger2017_subtypes}
{Taubenberger}, S. 2017, {The Extremes of Thermonuclear Supernovae}, ed. A.~W.
  {Alsabti} \& P.~{Murdin}, 317

\bibitem[{{Taubenberger} {et~al.}(2013){Taubenberger}, {Kromer}, {Pakmor},
  {Pignata}, {Maeda}, {Hachinger}, {Leibundgut}, \&
  {Hillebrandt}}]{Taubenberger2013_nebular}
{Taubenberger}, S., {Kromer}, M., {Pakmor}, R., {et~al.} 2013, \apjl, 775, L43

\bibitem[{{Thomas} {et~al.}(2011{\natexlab{a}}){Thomas}, {Nugent}, \&
  {Meza}}]{Thomas2011_synapps}
{Thomas}, R.~C., {Nugent}, P.~E., \& {Meza}, J.~C. 2011{\natexlab{a}}, \pasp,
  123, 237

\bibitem[{{Thomas} {et~al.}(2007){Thomas}, {Aldering}, {Antilogus}, {Aragon},
  {Bailey}, {Baltay}, {Baron}, {Bauer}, {Buton}, {Bongard}, {Copin}, {Gangler},
  {Gilles}, {Kessler}, {Loken}, {Nugent}, {Pain}, {Parrent}, {P{\'e}contal},
  {Pereira}, {Perlmutter}, {Rabinowitz}, {Rigaudier}, {Runge}, {Scalzo},
  {Smadja}, {Wang}, {Weaver}, \& {Nearby Supernova
  Factory}}]{Thomas2007_carbon}
{Thomas}, R.~C., {Aldering}, G., {Antilogus}, P., {et~al.} 2007, \apjl, 654,
  L53

\bibitem[{{Thomas} {et~al.}(2011{\natexlab{b}}){Thomas}, {Aldering},
  {Antilogus}, {Aragon}, {Bailey}, {Baltay}, {Bongard}, {Buton}, {Canto},
  {Childress}, {Chotard}, {Copin}, {Fakhouri}, {Gangler}, {Hsiao},
  {Kerschhaggl}, {Kowalski}, {Loken}, {Nugent}, {Paech}, {Pain}, {Pecontal},
  {Pereira}, {Perlmutter}, {Rabinowitz}, {Rigault}, {Rubin}, {Runge}, {Scalzo},
  {Smadja}, {Tao}, {Weaver}, {Wu}, {Brown}, {Milne}, \& {Nearby Supernova
  Factory}}]{Thomas2011_carbon}
---. 2011{\natexlab{b}}, \apj, 743, 27

\bibitem[{{Vink{\'o}} {et~al.}(2012){Vink{\'o}}, {S{\'a}rneczky}, {Tak{\'a}ts},
  {Marion}, {Heged{\"u}s}, {B{\'{\i}}r{\'o}}, {Borkovits}, {Szegedi-Elek},
  {Farkas}, {Klagyivik}, {Kiss}, {Kov{\'a}cs}, {P{\'a}l}, {Szak{\'a}ts},
  {Szalai}, {Szalai}, {Szatm{\'a}ry}, {Szing}, {Vida}, \&
  {Wheeler}}]{Vinko2012_11fe}
{Vink{\'o}}, J., {S{\'a}rneczky}, K., {Tak{\'a}ts}, K., {et~al.} 2012, \aap,
  546, A12

\bibitem[{{Wheeler} {et~al.}(1998){Wheeler}, {H{\"o}flich}, {Harkness}, \&
  {Spyromilio}}]{Wheeler1998_nir}
{Wheeler}, J.~C., {H{\"o}flich}, P., {Harkness}, R.~P., \& {Spyromilio}, J.
  1998, \apj, 496, 908

\end{thebibliography}

\appendix

\section{Outer zone}
\label{sec:appendix}

As argued in \S \ref{sec:trough}, carbon in the outer parts of the ejecta experience a lower density and temperature, meaning that under the NLTE approximations in TARDIS, the excitation levels that contribute to forming the optical feature are less likely to be populated. However, this effect may be compensated by greatly increasing the mass fraction of carbon in these layers. Here, we investigate the possible contribution from carbon at relatively high velocities by adding another zone to our analysis, where the mass fraction of carbon (\xco) is varied in a region defined by $16000 \lesssim v \lesssim 19000{\rm\,km\,s^{-1}}$. For reference, M14 adopts $\xc \gtrsim 0.02$ at comparable velocities.

As in \S \ref{subsec:plateaus}, we perform our analysis pair-wise, i.e., we investigate the acceptable mass fractions in the outer region, while allowing the carbon mass fraction in the middle region to vary. These zones are best constrained by the early spectra, for which the photosphere is the closest to these zones and at a time of minimal dilation. Note that the inner region is basically below the photosphere at $t_{\rm exp}$=3.7\,d and thus of less importance here.

\begin{figure*}
\epsscale{1.0}
\plotone{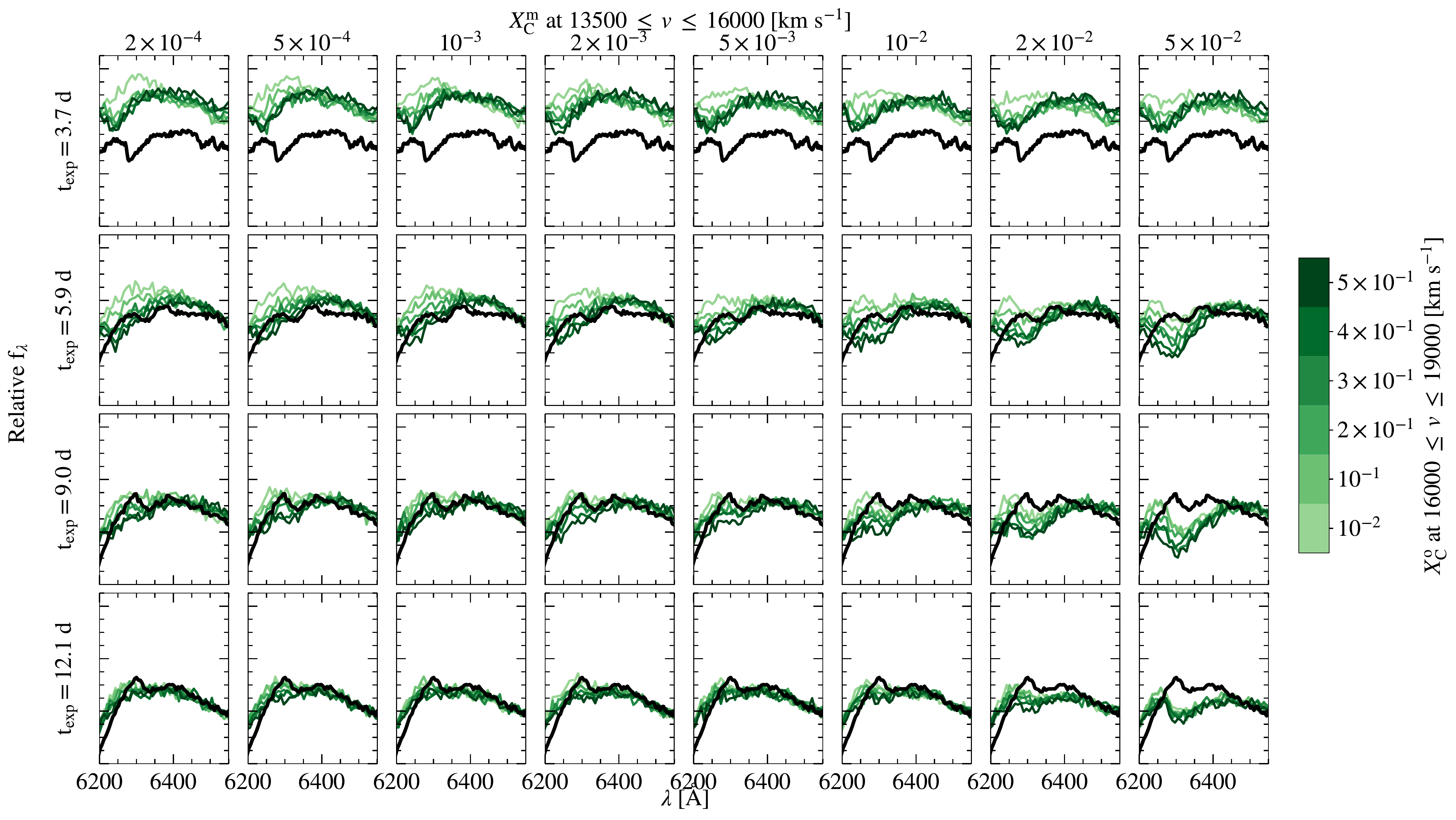}
\caption{Comparison of observed (black) and simulated (color) carbon profiles for a range of carbon abundances. Rows correspond to different epochs, while columns map the carbon mass fraction \xcm\ adopted for the middle region (velocity range $13500 \lesssim v \lesssim 16000{\rm\,km\,s^{-1}}$), and colors map the mass fraction \xco\ in the outer region ($16000 \lesssim v \lesssim 19000{\rm\,km\,s^{-1}}$). One sees, for instance, that models with $\xco \geq 0.4$ cannot explain the data at $t_{\rm exp}$=5.9, 9.0 and 12.1\,d for any choice of \xcm. $t_{\rm exp}$=16.1 and 19.1\,d are less affected by the middle and outer zones, as can be already seen at $t_{\rm exp}$=12.1\,d, and were not included for conciseness.}
\label{Fig:early_constraints}
\end{figure*}

In Fig. \ref{Fig:early_constraints} we show the synthetic spectra for which $0.01 \leq \xco \leq 0.5$ (color coded) and $0.0002 \leq \xcm \leq 0.05$. This range covers both the extremes of a carbon feature not forming at all in the early epochs, or the feature being either too strong or too blue. Here one has to be careful, because similarly to the spectra in M14, the red end of the main Si $\lambda$6355 is inconsistent with the data and will make a precise assessment of the blue end of the carbon feature less reliable. Correcting this discrepancy is beyond the scope of this work. Unfortunately, it also means that we can only place a conservative upper limit on \xco\ and lower limit on \xcm.

We conclude that $\xco\ \geq 0.4$ leads to a carbon trough that is inconsistent with the data at $t_{\rm exp}$=3.7, 5.9 and 12.1\,d for any choice of \xcm. Assessing a lower limit here is also difficult, given that none of the simulations seem to reproduce particularly well the spectra $t_{\rm exp}$=3.7\,d.

\end{document}